\documentclass[3p,twocolumn]{elsarticle}

\usepackage{amsmath,amssymb,amsfonts}
\usepackage{algorithmic}
\usepackage{graphicx}
\usepackage[inline]{enumitem}
\usepackage{pifont}
\usepackage{array}
\usepackage{lineno,hyperref}
\modulolinenumbers[5]
\usepackage[dvipsnames]{xcolor}
\usepackage{textcomp}
\usepackage{soulutf8} 
\modulolinenumbers[5]
\hypersetup{
    colorlinks=false,
    pdfborder={0 0 0},
}
\biboptions{sort&compress}

\newcommand{\cmark}{\ding{51}}%
\newcommand{\xmark}{\ding{55}}%

\newcommand{\refs}[1]{Section~\ref{#1}}
\newcommand{\reff}[1]{Figure~\ref{#1}}
\newcommand{\reft}[1]{Table~\ref{#1}}
\newcommand{\refe}[1]{Equation~\ref{#1}}

\journal{Journal of \LaTeX\ Templates}

\begin{document}

\begin{frontmatter}

\title{MobiGyges: A Mobile Hidden Volume for Preventing Data Loss,\\ Improving Storage Utilization, and Avoiding Device Reboot}

\author[1]{Wendi Feng}

\author[1]{Chuanchang Liu}

\author[2,3]{Zehua Guo}

\author[4]{Thar Baker}

\author[2,3]{\\ Gang Wang}

\author[1]{Meng Wang}

\author[1]{Bo Cheng}
\author[1]{and Junliang Chen}

\address[1]{Beijing University of Posts and Telecommunications, 10 Xitucheng RD, 100876, Beijing, China}
\address[2]{Beijing Institute of Technology, 5 Zhongguancun ST South, 100081, Beijing, China}
\address[3]{University of Minnesota Twin Cities, 117 Pleasant ST, 55455, Minneapolis, USA}
\address[4]{Liverpool John Moores University, James Parson Building, Liverpool, L3 3AF, UK}

\begin{abstract}

Sensitive data protection is essential for mobile users. Plausibly Deniable Encryption (PDE) systems provide an effective manner to protect sensitive data by hiding them on the device. However, existing PDE systems can lose data due to overriding the hidden volume, waste physical storage because of the ``reserved area'' used for avoiding data loss, and require device reboot when using the hidden volume. This paper presents MobiGyges, a hidden volume based mobile PDE system, to fill the gap. MobiGyges addresses the problem of data loss by restricting each storage block used only by one volume, and it improves storage utilization by eliminating the ``reserved area''. MobiGyges can also avoid device reboot by mounting the hidden volume dynamically on-demand with the \textit{Dynamic Mounting} service. Moreover, we identify two novel PDE oriented attacks, the \textit{capacity comparison attack} and the \textit{fill-to-full attack}. MobiGyges can defend them by jointly leveraging the \textit{Shrunk U-disk method} and \textit{multi-level deniability}. We implement the MobiGyges proof-of-concept system on a real mobile phone Google Nexus 6P with LineageOS 13. Experimental results show that MobiGyges prevents data loss, avoids device reboot, improves storage utilization by over 30\% with acceptable performance overhead compared with current works.

\end{abstract}


\end{frontmatter}


%
\section{Introduction}
\label{sec:intro}
Mobile devices (e.g., smartphones) have become prevalent in recent years, especially in the era of 5G and the Internet of Things (IoTs)~\cite{gangiot}. Hence, protecting private and sensitive data on mobile devices are very important to users~\cite{iont,beamcom}. One existing solution is to use Full Disk Encryption (FDE)~\cite{gotzfried2014analysing}. FDE uses an encrypting key to encrypt user data before storing it on a device and decrypt the data before applications using it~\cite{androidfde}. Nonetheless, FDE is not secure because sensitive data can be compromised when the encryption key is exposed, since the encrypted data can be easily decrypted with the only encryption key.

Recent works~\cite{canetti1997deniable,skillen2014mobiflage,yu2014mobihydra,chang2015mobipluto,userfriendly} propose Plausibly Deniable Encryption (PDE) to enhance the security. PDE is a data protection paradigm that protects sensitive data on both stationary systems and mobile systems by providing \textit{deniability} for the sensitive data. Deniability means that sensitive data owner can deny the existence of the data. Modern PDE systems use the \textit{hidden volume mechanism} to implement the deniability. The hidden volume based PDE system stores the sensitive data on the hidden volume, yet the hidden volume itself is concealed inside the device. Logically, the storage space on the hidden volume based PDE systems can be divided into \textit{hidden volumes} and an \textit{outer volume}. The outer volume is visible to all users for daily purposes and will be used automatically as the system starts up, while hidden volumes are concealed in the device and store the sensitive data.

\begin{figure}[t]
\centering
\includegraphics[width=0.9\columnwidth]{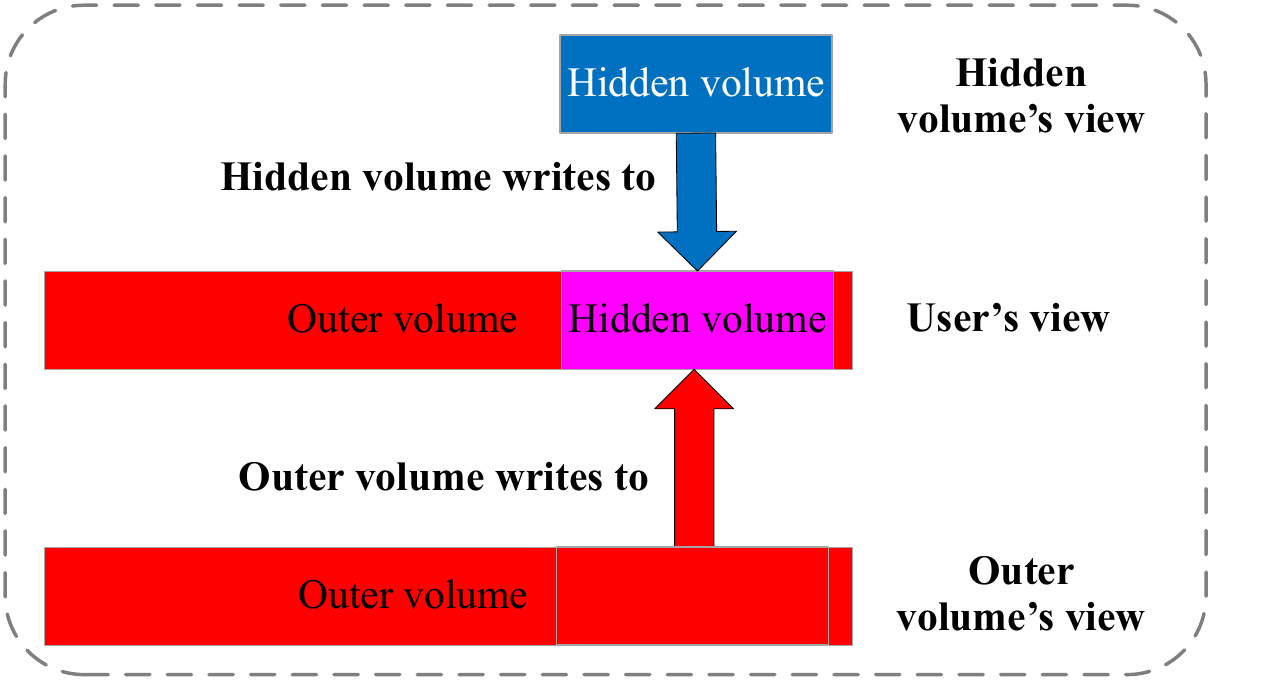}
\caption{Data loss caused by data override. The outer volume writes data on hidden volume occupied blocks.}
\label{fig:override}
\end{figure}

\begin{figure}[t]
\centering
\includegraphics[width=0.9\columnwidth]{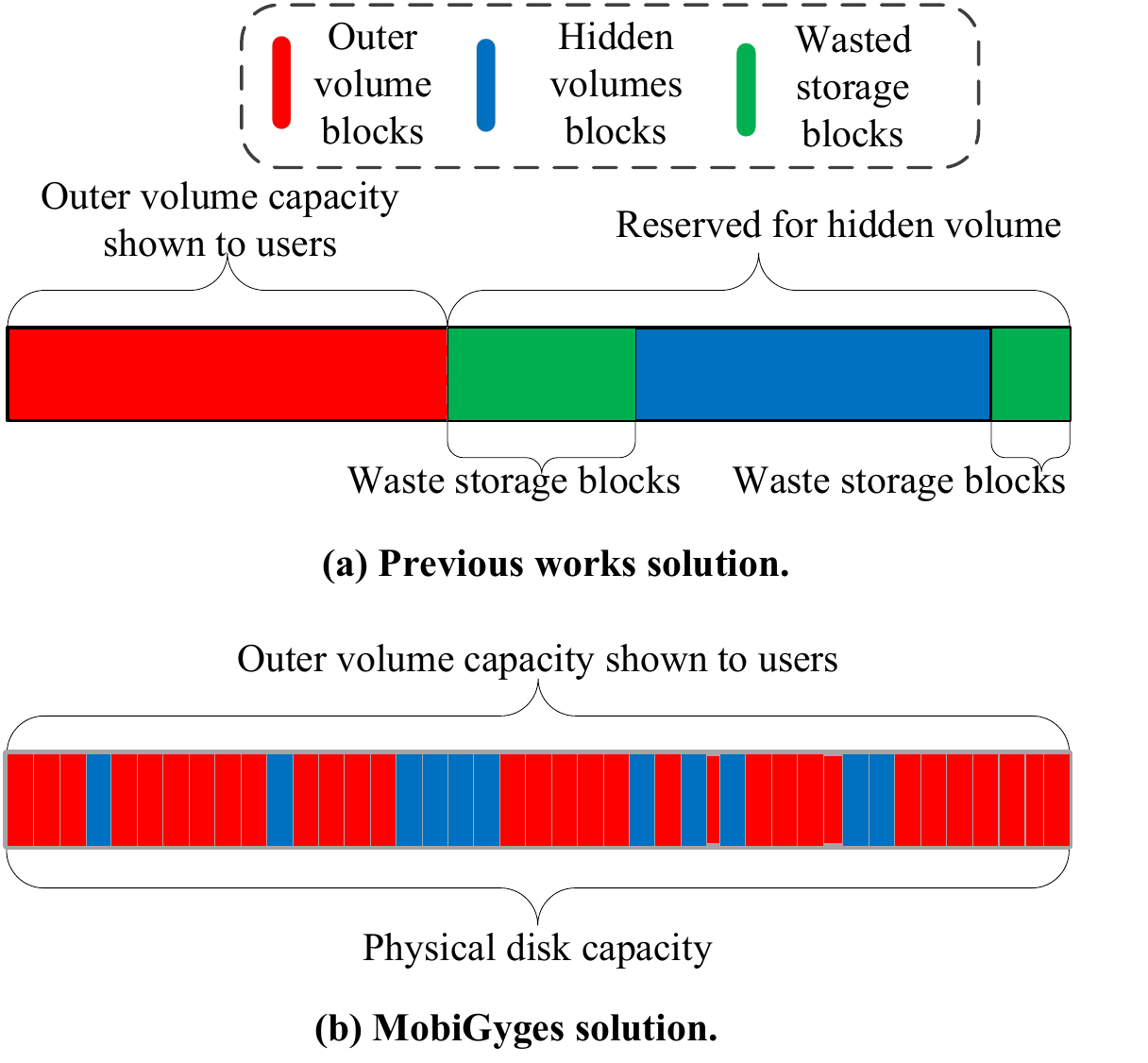}
\caption{Hidden volume is placed into a reserved area to avoid data override. Large amount of storage space is wasted. MobiGyges can fully utilize almost all the storage space.}
\label{fig:waste}
\end{figure}

Such hidden volume based solutions have the following limitations:
\begin{itemize}

\item \textbf{Data loss.} Hidden volumes are concealed inside the outer volume~\cite{skillen2014mobiflage,yu2014mobihydra,chang2015mobipluto,userfriendly,hong2017personal}, but the outer volume does not know the existence of hidden volumes. Therefore, it is likely to write data on the storage blocks that are occupied by hidden volumes. Thus, sensitive data stored on the hidden volume will be lost. \reff{fig:override} shows an example of data overriding between the outer volume and the hidden volume. In the figure, the outer volume considers all the storage space ({\color{red}{red space}}) to be usable. When the outer volume writes data on the space of the hidden volume ({\color{Thistle}{purple space}}), the data on the hidden volume will be lost.

\item \textbf{Storage waste.} Studies~\cite{skillen2014mobiflage,yu2014mobihydra,chang2015mobipluto,userfriendly,mobiceal} attempt to solve the data loss problem by placing the hidden volume into a ``reserved area'', and the outer volume will not write data to that area. The size of the ``reserved area'' is bigger than the capacity of the hidden volume. As depicted in \reff{fig:waste}a, in previous works, the right part ({\color{ForestGreen}{green blocks}} plus the {\color{RoyalBlue}{blue block}}) of the physical volume is reserved for the hidden volume. Since the capacity of the hidden volume (the {\color{RoyalBlue}{blue block}}) is much smaller than that of the ``reserved area'', the hidden volume can ``float'' inside the ``reserved area''. Hence, the exact starting position of the hidden volume can be arbitrary, hidden volumes are thus protected. However, this mechanism could waste a large amount of storage space ({\color{ForestGreen}{green blocks}}). We find in these solutions~\cite{skillen2014mobiflage,yu2014mobihydra,chang2015mobipluto}, up to 45\%\footnote{Including storage space taken up by the structure of a file system. The calculation of the utilization is detailed in \refs{sec:evalueate}.} of the total storage space is wasted, which is huge for the resource-limited mobile devices.

\item \textbf{Device reboot.} State-of-the-art works use two modes in their system, namely, the \textit{normal mode} and the \textit{PDE mode}~\cite{skillen2014mobiflage,yu2014mobihydra,chang2015mobipluto,userfriendly,hong2017personal}. The normal mode uses the outer volume while the PDE mode uses the hidden volume, respectively. When users want to use the hidden volume, they have to use the PDE mode. However, device reboot is required to switch modes. Rebooting the device to use the PDE mode wastes time and is not convenient for users especially those who want to use the hidden volume urgently.

\end{itemize}

Apart from the drawbacks, we identify two possible PDE oriented attacks (detailed in \refs{sec:attacks}) that might compromise the sensitive data, and current solutions fail to defend.

\begin{itemize}

\item \textbf{The capacity comparison attack.} The attacker may discover the hidden volume by comparing the capacity of the outer volume and the hidden volume. If their capacities are different, the attacker can doubt the device is particularly designed, which is prone to expose the hidden volume and hence compromises the sensitive data. For example, a 32GB device uses 5GB for the hidden volume, so the capacity of the outer volume is 27GB. The attacker can doubt about the 5GB capacity difference, and conduct further investigation.

\item \textbf{The fill-to-full attack.} If the attacker identifies the potential existence of the hidden volume, he/she may conduct the fill-to-full attack to explore the real capacity of the outer volume by writing arbitrary data to the outer volume and filling it until full. After filling the outer volume, the attacker gets the audited information and conducts the capacity comparison attack. If the real capacity is different from that of the physical disk, the hidden volume will be compromised.

\end{itemize}

Existing solutions cannot solve the three problems simultaneously, and they cannot defend the two attacks. To this end, we present MobiGyges in this paper. MobiGyges is a hidden volume based PDE system. It introduces the Volume Management module and FDE module, which prevents sensitive data loss by restricting each storage block usable by solely one volume, and improves the storage utilization by eliminating the ``reserved area'' (as shown in \reff{fig:waste}b), and avoids rebooting the device to use the hidden volume by introducing the Dynamic Mounting service that mounts the hidden volume on-demand. MobiGyges also uses the \textit{Shrunk U-disk} method (detailed in \refs{sec:functionality}\ref{sec:cap_comp}\textsl{}) and \textit{multi-level deniability} (detailed in \refs{sec:composition}\ref{sec:multideniability}) to jointly defend the aforementioned attacks. 

\begin{figure}[t]
\centering
\includegraphics[width=\columnwidth]{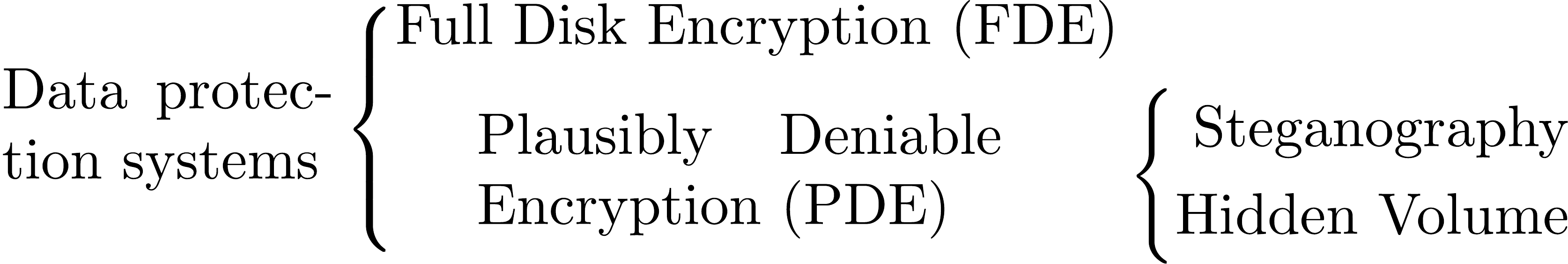}
\caption{Data protection systems classification.}
\label{fig:systems}
\end{figure}


Our main contribution in this paper is threefold, summarized as follows.

\begin{itemize}
\item We propose MobiGyges to solve the data overriding problem, improve storage utilization with the aid of Thin Provisioning and Device Mapper, and avoid device reboot to use the hidden volume on-demand by introducing the Dynamic Mounting service.

\item We identify the capacity comparison attack and the fill-to-full attack, and propose the Shrunk U-disk method and multi-level deniability to jointly defend them.

\item We implement the MobiGyges proof-of-concept system on Google Nexus 6P with the LineageOS~\cite{lineageos} 13 by porting Thin-Provisioning (pdata\_tools) and Logical Volume Management (LVM) into the Android system and implementing a TriggerApp to use hidden volume on-demand secretly. We conduct experiments to evaluate MobiGyges's storage utilization, performance overhead, and experimental results show that MobiGyes reaches all our design goal and improves storage utilization by over 30\% compared with current solutions.

\end{itemize}

The rest of the paper is organized as follows. Section~\ref{sec:works} introduces related works, and Section~\ref{sec:threat_model} presents the threat model and assumptions. In Section~\ref{sec:attacks}, we introduce our newly identified PDE oriented attacks. Section~\ref{sec:design} presents the design of MobiGyges. In Section~\ref{sec:impl}, we present the implementation of MobiGyges with LineageOS 13 on Google Nexus 6P. In Section~\ref{sec:evalueate}, we conduct rigorous experiments and analyze the experimental results. 
Section~\mbox{\ref{sec:discussion}} discusses common attacks defended by MobiGyges, the drawback, and possible future works.
Finally, we conclude the paper in Section~\ref{sec:conclude}.

\section{Related Works}
\label{sec:works}

\begin{figure}[t]
\centering
\includegraphics[width=0.9\columnwidth]{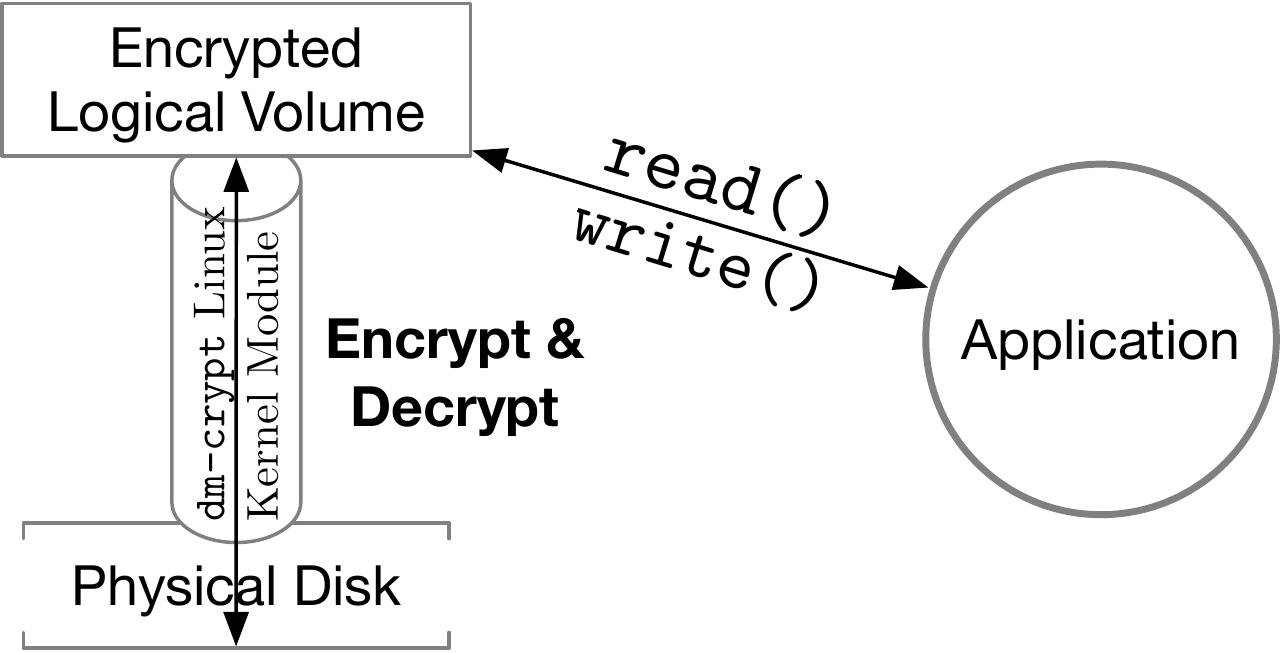}
\caption{Encrypted logical volume and physical storage. Applications use the normal system call to read or write data to the encrypted logical volume. The \texttt{dm-crypt} Linux module automatically encrypts and decrypts the data between the encrypted logical volume and the physical disk.}
\label{fig:fde_demo}
\end{figure}


Data protection is of paramount importance, and there have already been various systems proposed for security. In this section, we categorize current data protection systems (as shown in \reff{fig:systems}) and introduce related works based on the taxonomy. Other types of file systems like versioning file system~\cite{trustedfs} that are useful in post-intrusion file system analysis applications, or reliable file retention and retrievability required by legal regulations for sensitive data management. However, they are not suitable for personal sensitive data protection. Hence, they are out of the scope of this paper and, therefore, not discussed here.

\subsection{Full Disk Encryption}

FDE is an elementary way to prevent sensitive data from attacks by encrypting all data on a volume. As shown in Figure \ref{fig:fde_demo}, on Linux based systems, FDE creates an encrypted logical volume with \texttt{dm-crypt} Linux kernel module, and it encrypts all the data that saved on the logical volume before committing to the physical storage. Similarly, when an application needs to read data, it automatically decrypts the data from the physical storage and redirects the data to the logical volume. Thus, FDE is transparent to applications. FDE has now been a standard configuration for any security system. There are many mature FDE solutions, e.g., TrueCrypt~\cite{team18truecrypt,czeskis2008defeating}, BitLocker~\cite{bitlocker,kumar2008bitlocker}, LUKS (Linux Unified Key Setup)~\cite{fruhwirth2009luks}, FileVault~\cite{os2014filevault}.

TrueCrypt \cite{team18truecrypt,czeskis2008defeating} is a cross-platform encryption software that supports multiple cipher encryption scheme. It underpins various ways for both software architecturally and hardware chip performance improvement to speed up the full disk encryption process. TrueCrypt supports PDE, which will be introduced in \refs{sec:related_hidden}.

BitLocker \cite{bitlocker,kumar2008bitlocker} is a built-in encryption utility software on Windows developed by Microsoft to conduct FDE. It has been a system-level component since Windows Vista. The default encryption scheme is AES with cipher block chaining (CBC) or XTS with a 128-bit or 256-bit key. Note that CBC is not used over the whole disk but applied to each individual sector.

LUKS \cite{fruhwirth2009luks} (Linux Unified Key Setup) is an encryption specification for Linux used for employing FDE. Unlike most of the encryption software creates their own encryption functionality. LUKS creates a unified encryption format that can be used for various tools like \texttt{cryptsetup}.

FileVault \cite{os2014filevault} is another FDE solution introduced with Mac OS X Panther by Apple Inc. As recommended by NIST \cite{dworkin2010recommendation}, it uses the AES-XTS mode of AES with 128-bit blocks and a 256-bit key to encrypt the disk.

However, since FDE uses only one encryption key, all the FDE-only solutions fail to provide users the deniability of sensitive data stored on their devices, which is not enough for protecting the sensitive data.

\subsection{Plausibly Deniable Encryption}
PDE is the kind of encryption paradigm that provides the user with the ability to deny the existence of sensitive data on the device. There are primarily two ways of implementing PDE: Steganography and the hidden volume.

\begin{table*}[t]
\centering
\caption{Feature comparison between MobiGyges and current works. (\cmark\ means the functionality is provided, \xmark\ means the functionality is not available, and \textemdash{} means not applicable.)}
\label{tbl:functional}
\resizebox{!}{2cm}{
\begin{tabular}{|m{0.175\textwidth}|c|c|c|c|c|c|}
\hline
\textbf{Features} & \textbf{Mobiflage \cite{skillen2014mobiflage}} & \textbf{MobiHydra \cite{yu2014mobihydra}} & \textbf{MobiPluto \cite{chang2015mobipluto}} & \textbf{MobiMimosa \cite{hong2017personal}} & \textbf{MobiCeal \cite{mobiceal}} & \textbf{MobiGyges} \\ \hline \hline
Data loss prevention & \cmark  & \cmark & \cmark & \xmark & \cmark & \cmark \\ \hline
Reserved area elimination & \xmark  & \xmark & \xmark & \textemdash & \xmark & \cmark \\ \hline
Device Reboot avoidance & \xmark & \xmark & \xmark & \xmark & \cmark & \cmark \\ \hline
Capacity comparison attack defense & \xmark & \xmark & \xmark & \xmark & \xmark & \cmark \\ \hline
Fill-to-full attack defense & \textemdash & \textemdash & \textemdash & \textemdash & \textemdash & \cmark \\ \hline
\end{tabular}
}
\end{table*}

\subsubsection{Steganography}

An example of Steganography is hiding sensitive data into a multimedia file like a photo, a video or an audio file as noise points. Since people tend to ignore those noise points, the sensitive data is thus protected. However, this technique requires a large amount of computation, which is not appropriate for mobile devices, because mobile devices lack of computation and storage resources. StegFS~\cite{mcdonald1999stegfs} is a steganography-based file system, and its key idea is to hide data in a bunch of cover files. Another work~\cite{artifice} uses external entropy sources and erasure codes to deniably and reliably store data within the unallocated space of an existing file system. However, these solutions has the following shortcomings.
\begin{enumerate*}[label=\roman*)]
\item It wastes storage space. 
\item The performance is low especially when writing.
\item The possibility of data loss is high. 
\item The modification of Ext2 may lead to compromise of deniability.
\end{enumerate*}
All these shortcomings make it not suitable for mobile devices.

\subsubsection{Hidden volume} 
\label{sec:related_hidden}

Hidden volumes achieve PDE by concealing themselves into a device. It is light weighted and has minimum computational and storage burdens. However, current hidden volume solutions have mainly three drawbacks that we have pointed out in \refs{sec:intro}, and these drawbacks cannot be addressed at the same time. In this section, we classify works based on the drawbacks they solve as follows.

\begin{enumerate}[label=\textbf{(\alph*)}, wide]

\item\textbf{Data loss reducing solutions}

TrueCrypt~\cite{team18truecrypt} and FreeOTFE~\cite{freeotfe} are PC PDE solutions. They can create hidden volume(s) as files or physical volumes. However, both TrueCrypt and FreeOTFE can only create PDE hidden volume on their resource files. Therefore, if these files are broken or lost, all the data stored on the hidden volumes will be lost. Thus, the solution is prone to compromise the sensitive data.

Mobiflage~\cite{skillen2014mobiflage} is the first implementation of mobile hidden volume based PDE prototype system on Android. To avoid sensitive data loss, Mobiflage reserves a block of storage space and places the hidden volume at an arbitrary position in the area. We call it the ``reserved area'' technique. However, up to 45\% of the total physical storage is wasted. Based on Mobiflage, MobiHydra~\cite{yu2014mobihydra} implements multi-level deniability. However, similar to Mobiflage, MobiHydra also fails to solve the problem of low storage utilization.

MobiPluto~\cite{chang2015mobipluto,userfriendly} is the first file system friendly PDE solution built on the Android operating system. It leverages virtual logical volumes techniques, and all block-based file system can run on it without modifications. It also addresses the data loss problem by placing the hidden volume into a ``reserved area'', therefore, the storage utilization is low.

MobiCeal~\cite{mobiceal} is another recent hidden volume PDE solution, and its key contribution is to defend against strong coercive multi-snapshot adversaries. It does not consider the storage waste and capacity inconsistency problems in previous solutions. 

\item\textbf{High storage utilization solutions}

MobiMimosa~\cite{hong2017personal} is our former work of the hidden volume based PDE solution for the Android. Its key idea is manually choosing storage blocks on a physical device and converting the blocks into a hidden volume. However, the location information of the storage blocks is stored on a \texttt{dm\_table} file. Once getting the file, the attacker can easily get the sensitive data, and if the \texttt{dm\_table} file is lost, all the sensitive data on the hidden volume will be lost.

TrustGyges~\cite{trustgyges} addresses the above problem by storing the \texttt{dm\_table} file on to a cloud server. In order to avoid run-time attack, it fetches the \texttt{dm\_table} file inside the Trust Execution Environment (TEE). However, it fails to propose a proper method to use the hidden volume without the network connection.

Our earlier work~\cite{mobigyges_iscc} first proposes to use the Shrunk U-disk method to hide the hidden volume, which achieves high storage utilization and avoids data loss simultaneously. However, it fails to support using the hidden volume without rebooting the device, and it cannot mitigate the fill-to-full attack.

\item \textbf{Reboot avoiding solutions}

MobiCeal~\cite{mobiceal} is the latest PDE solution and is the only mobile PDE solution that does not require device reboot. Similar to all other solutions, it still has the concept of normal mode and PDE mode. Its main concern is how to fastly switch from the public mode to the hidden mode (PDE mode). It achieves the device reboot avoidance by restarting the Android framework. Although, it reduces the switch time, but the time of restarting the Android framework is still too long ($\approx$10s for entering the PDE mode and $\approx$70s for returning to normal mode). Furthermore, applications will suffer response lags after restarting the Android framework that results from CPU cache misses, and hence influences the user experience.

\end{enumerate}

\subsubsection{Functionality Comparison}
\label{sec:functional}
Although mobile PDE has been explored by~\cite{skillen2014mobiflage,yu2014mobihydra,chang2015mobipluto,hong2017personal,mobiceal}, we differentiate MobiGyges with them in \reft{tbl:functional} on key features desired for PDE oriented data protections.

First, most of the current solutions use the ``reserved area'' to prevent data loss resulting from data override on the hidden volume. Therefore, they fail to utilize the storage space efficiently. Second, all current works cannot defend our newly identified two novel PDE oriented attacks. Consequently, the possibility of exposing the hidden volume and compromising the sensitive data is high. Our proposed MobiGyges system can address all of these issues at the same time.

\section{Threat Model and System Assumptions}
\label{sec:threat_model}

The key to protect sensitive data on a hidden volume based PDE system is concealing the hidden volume~\cite{jia2017deftl}. We propose the following threat model and put the beneath assumptions for MobiGyges based on works presented in~\cite{skillen2014mobiflage, yu2014mobihydra}.
\begin{enumerate}
\item The attacker knows that the \texttt{userdata} partition is encrypted by FDE and also the key of this encryption. But the attacker lacks the knowledge of MobiGyges’s design. Thus, the attackers still cannot retrieve the sensitive data because they do not know where and how to do that.  Alternatively, the attacker knows about the design of MobiGyges; but, not sure how many hidden volumes there are on the phone.
\item The attacker knows the design of MobiGyges but is uncertain if the key provided by users is the key to the hidden volume he sought.
\item The attacker can get root privilege and the physical storage of the phone or dump the raw data from the storage medium.
\end{enumerate}
It is notoriously hard to achieve security on a device with backdoor hardware or software. Therefore, assumptions involving the backdoors of the device are necessary for us to design MobiGyges. We make the following assumptions accordingly.
\begin{enumerate}
\item The hardware of users’ phone is backdoor-free.
\item The system level software (e.g., boot-loader and the mobile operating system) are backdoor-free.
\end{enumerate}
Without these assumptions, user's operational behavior can be monitored, which is impossible to conceal the hidden volume and protect the sensitive data stored on it.

\section{Novel PDE Oriented Attacks}
\label{sec:attacks}

\begin{table}[t!]
\centering
\caption{Notation definitions used throughout the paper}.
\label{tbl:notation}
\begin{tabular}{l|p{0.7\columnwidth}}
\hline
\textbf{Notation} & \textbf{Description} \\ \hline 
\hline
$\mathcal{A}$ & An attacker. \\\hline
$D$ & A mobile device. \\\hline
$C_p$ & The physical capacity of the device $D$.\\ \hline
$C_o$ & The capacity of the device $D$'s outer volume.\\\hline
$O_l$ & The sector offset of the mapping logical volume.\\\hline
$S_l$ & The number of sectors of the original volume.\\\hline
$T$ & The mapping type.\\\hline
$D_p$ & The mapped device. \\\hline
$O_p$ & The offset of the mapped physical volume.\\\hline
$S_m$ & The size of metadata volume.\\\hline
$S_p$ & The size of pool volume. \\\hline
$S_c$ & The chunk size of pool volume.\\\hline
$N$ & The name of the hidden volume.\\\hline
$T(s, b)$ & The trim function that trims string $s$ into a $b$-length string.\\\hline
$h(x)$ & The hash function that hashes variable $x$ into a hash value.\\\hline
$\eta$ & The storage utilization.\\\hline
\textit{Offset} & The hidden volume offset in Mobiflage~\cite{skillen2014mobiflage}.\\\hline
\textit{vlen} & The capacity of the device in Mobiflage~\cite{skillen2014mobiflage}.\\\hline
$H(x)$ & The PBKDF2 iterated hash function in Mobiflage~\cite{skillen2014mobiflage}.\\\hline
\textit{pwd} & The password of the hidden volume in Mobiflage~\cite{skillen2014mobiflage}.\\\hline
\textit{salt} & The random salt value for PBKDF2 in Mobiflage~\cite{skillen2014mobiflage}.\\\hline
\end{tabular}
\end{table}

In this section, we introduce our identified two novel PDE oriented attacks and assume attacker $\mathcal{A}$ conducts the following attacks. Since these two attacks are correlated with each other, we propose joint solutions to defend the attacks.

\subsection{Capacity Comparison Attack}
\label{sec:compare_attack}
When PDE systems are created using hidden volumes, the total capacity of the outer volume and hidden volumes should be equal to the physical capacity. Hence, if untreated, the capacity of the out volume will be smaller than the physical capacity. By comparing the capacities of the outer volume and the physical disk, the attacker can get the capacity inconsistency, and the capacity inconsistency offers attacker $\mathcal{A}$ an indication of the potential existence of hidden volumes, which may lead to the attacker $\mathcal{A}$ to conduct a further investigation. Thus, the attack makes hidden volume based PDE system highly prone to compromise the sensitive data. We formalize the attack as shown in Equation~\ref{eq:cca}, and all notation definitions can be found in Table~\ref{tbl:notation}.
\begin{equation}
\label{eq:cca}
\mathcal{A}\leftarrow D \leftarrow\left \{
\begin{aligned}
1,& &C_p > C_o\\
0,& &C_p = C_o\\
\end{aligned}
\right .,
\end{equation}
where, $D$ denotes the device, $C_p$ is the physical capacity of the device, and $C_o$ is the capacity of the outer volume. 1 represents device $D$ has special design of its storage system (PDE is compromised), and 0 otherwise.

Existing literature fails to defend the capacity comparison attack. According to \refe{eq:cca}, we can leverage the second condition in the equation to defend the attack by eliminating the capacity difference between the outer volume and the physical disk. To this end, we have to find a method to modify the capacity of the outer volume $C_o$ to be the same as the physical capacity $C_p$. We detail the method used of defending the attack in \refs{sec:design}.

\subsection{Fill-to-Full Attack}
\label{sec:ftf_attack}
The fill-to-full attack is a complementary attack of the capacity comparison attack. Suppose now we can defend the capacity comparison attack by setting the capacity of the outer volume to be the same as that of the physical disk. Attacker $\mathcal{A}$ may still doubt that the device might have hidden volumes, but he/she is uncertain about it. To this end, attacker $\mathcal{A}$ may write arbitrary data on the outer volume by filling data to the outer volume and auditing the total size of the data that have been written. After getting the audited information, attacker $\mathcal{A}$ can further conduct the capacity comparison attack by comparing the physical capacity with the audited data size plus used capacity on the outer volume before the fill-to-full attack.

We demonstrate the fill-to-full attack with a real-world example. Suppose Alice has two containers: one is a standard 1-liter container, and another is marked as 3 liters. Alice wants to know if the second container is a 3-liter container precisely as it marked. She can fill up the 1-liter container with water and pour the water to the 3-liter container for 3 times. If the total amount of water cannot fill up the 3-liter container, or the 3-liter container overflows, Alice can confirm that the 3-liter container does not have a 3-liter capacity. The fill-to-full attack uses the same strategy.

Since the real size of the outer volume is smaller than the physical capacity. The audited size should be smaller than the physical capacity, which again reduces to the capacity comparison attack as we introduced in Section~\ref{sec:compare_attack}. In the next section, we introduce MobiGyges design, including how we defend the two attacks.

\section{MobiGyges Design}
\label{sec:design}

In this section, we introduce the design of MobiGyges. MobiGyges is designed for protecting sensitive user data rather than system data that include program executables and data. We first present our design considerations, and we then propose our design overview. As the Volume Management module is the key component of MobiGyges, we finally make a detailed anatomy of the design of the Volume Management module.

\subsection{Design Considerations}
\label{sub_sec:consider}
The current hidden volume based PDE solutions cannot address the aforementioned three problems at the same time, and they also fail to defend the capacity comparison and the fill-to-full attacks. MobiGyges is designed to conquer them all.

\begin{enumerate}[label=\textbf{(\arabic*)}, wide]

\item\textbf{Eliminating the data override issue.} The root of data loss is due to the data override phenomenon between the outer volume and the hidden volume(s), which two or more volumes access the same storage block. Therefore, if we restrict one physical storage block that can only be used by one volume, the issue is addressed. Thus we can split the total storage blocks into fine-grained storage blocks instead of coarsely divide the whole physical storage into two parts, and the outer volume and the hidden volume(s) allocate storage space from the fine-grained storage blocks (as shown in {\reff{fig:waste}b}).

\item \textbf{Improving the storage space utilization}. As we have introduced, the low storage utilization results from the ``reserved area''. Therefore, eliminating the ``reserved area'' can improve the storage utilization.

\item \textbf{Avoiding device reboot}. Rebooting the device costs time, which is not convenient for users, especially when users have to capture some important data and want to store it immediately into the hidden volume. We intend to avoid device reboot by mounting the hidden volume on-demand without rebooting the device.

\item \textbf{Defending the novel PDE oriented attacks.} The capacity comparison and the fill-to-full attacks result from the capacity inconsistency issue. The challenge is how we defend the fill-to-full and capacity comparison attacks.

\end{enumerate}

\begin{figure}[t]
\centering
\includegraphics[width=\columnwidth]{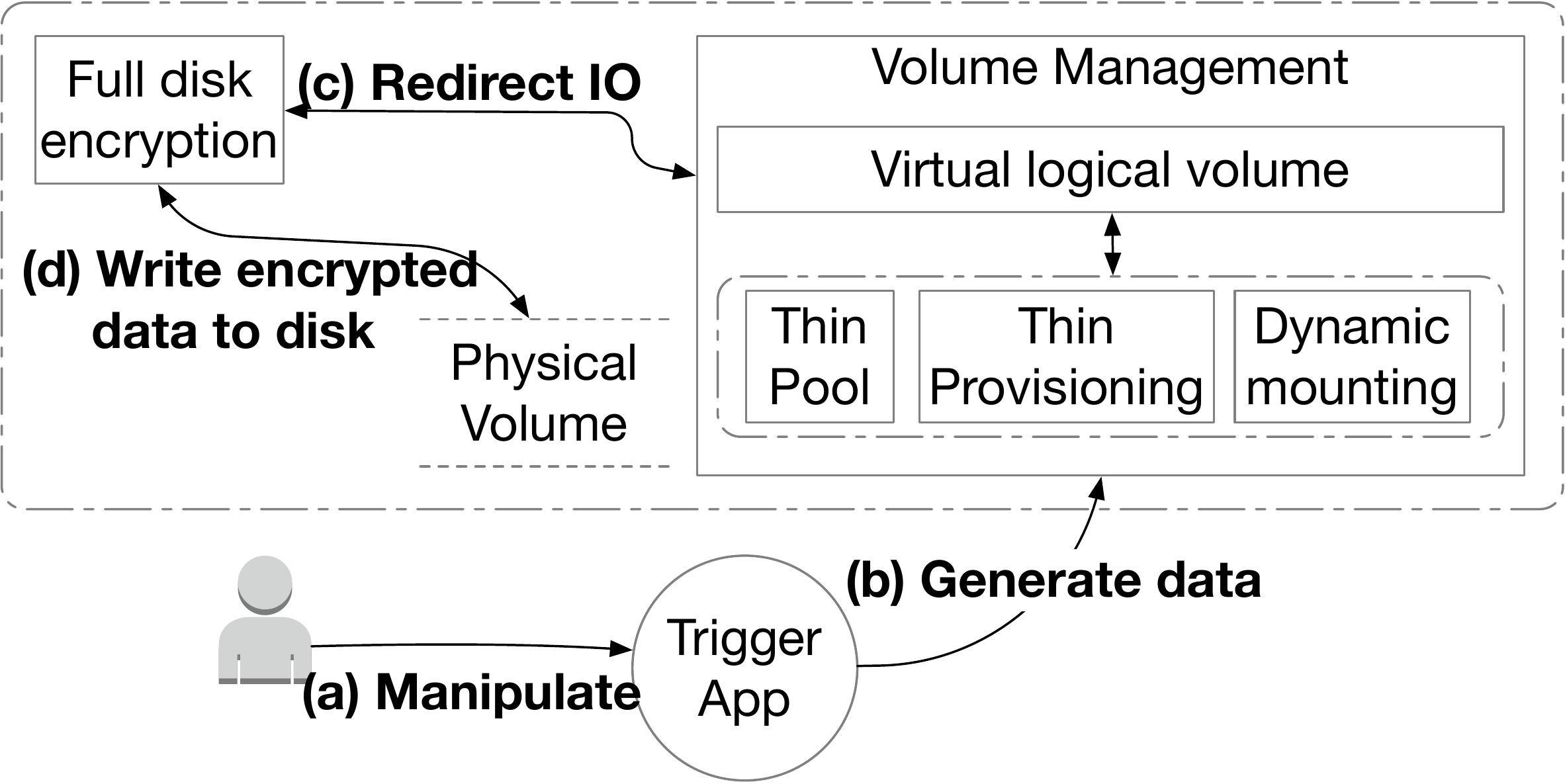}
\caption{MobiGyges key components and data-flow.}
\label{fig:overview}
\end{figure}


\subsection{Design Overview}
\label{sec:overview}

MobiGyges is a mobile hidden volume based PDE system. In this subsection, an overview of MobiGyges design, starting with a general description of its components, is presented. Then, we propose a so-called \textit{Shrunk U-Disk method} to defend the capacity comparison attack, and we also leverage the \textit{multi-level deniability} to defend the fill-to-full attack. Next, we introduce our design solution associated with the design consideration introduced in \refs{sub_sec:consider}. Finally, we describe the user steps.

\subsubsection{Solutions}

With the design considerations discussed in \refs{sub_sec:consider}, we propose our PDE system design solutions.
\begin{enumerate*}[label=\roman*)]
\item We introduce an independent component called the Volume Management module to manage the outer volume and hidden volume(s) rather than having the volumes themselves handle the hidden volume concealing. It flexibly allocates separate storage blocks for volumes and avoids the override of the same storage space. The module also eliminates the ``reserved area'' and thus mitigates data loss and improves storage utilization. We use a FDE component to encrypt the Volume Management module and the file system structure, which protects the system.

\item We design a userspace application called TriggerApp and the dynamic mounting service to mitigate device reboot. The dynamic mounting service can mount the hidden volume without rebooting the device, and TriggerApp can secretly trigger the use of hidden volumes and mount the hidden volume on-demand using the dynamic mounting service without rebooting the device.

\item We use the \textit{Shrunk U-disk method} and \textit{multi-level deniability} jointly in the Volume Management module to defend the capacity comparison and the fill-to-full attacks. In terms of fill-to-full attack, we leverage multi-level deniability by recording the size of data that has already been written and redirecting the extra attack IO requests to other places that will not take up the physical storage space and finally making the size of the attack IO plus the used capacity before the attack equals the physical capacity. Moreover, we employ the Shrunk U-disk method to defend the capacity comparison attack by intentionally labeling the capacity of the outer volume capacity equivalent to the physical capacity. We detail the Shrunk U-disk method and multi-deniability in \refs{sec:functionality}\ref{sec:cap_comp} and \refs{sec:composition}\ref{sec:multideniability}.
\end{enumerate*}

\begin{figure}[t]
\centering
\includegraphics[width=\columnwidth]{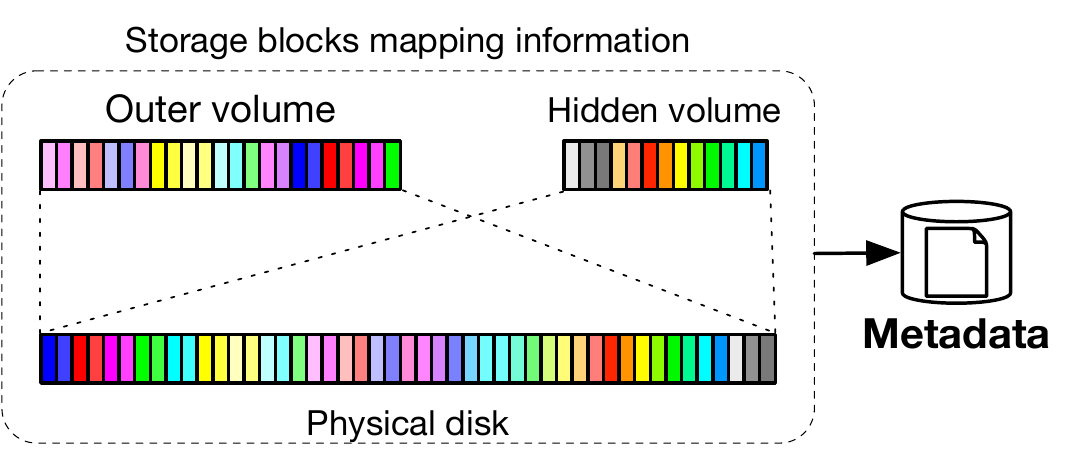}
\caption{Splitting the physical disk into fine-grained storage blocks (each block is represented by different colors), and each fine-grained block is used only for one volume (blocks of the outer volume and the hidden volume are allocated from the split physical disk). Thus, the ``reserved area'' is not required in the design. The occupancy information is stored into the metadata datastore.}
\label{fig:prob_avoid}
\end{figure}

   
\subsubsection{Key Components}

MobiGyges consists of four main components. \textit{Volume Management module, \textit{FDE module}, \textit{physical volume module}, and \textit{TriggerApps module}}.
\begin{enumerate*}[label=\roman*)]
\item The Volume Management module is the most important module of MobiGyges, and it manages the life-cycle of all the outer volume and hidden volumes.
\item FDE module is the encryption layer on top of the physical storage to protect the Volume Management module, and
\item physical volume module is the physical storage of the device provided by the device vendor.
\item TriggerApp module is an application, which is designed for the user to utilize the hidden volume securely and conveniently without requiring entering the detailed system commands.
\end{enumerate*}

\reff{fig:overview} depicts the data-flow in the entire MobiGyges workflow and the inner data-flow between MobiGyges components. As per the mentioned figure, 
\begin{enumerate*}[label=\textbf{(\alph*)}]
\item users first manipulate TriggerApps, and 
\item TiggerApps then generates data and exchanges data between the Volume Management. 
\item Next, the Volume Management module processes the data with \textit{Device Mapper} and \textit{Thin Provisioning}, and sends Input/Output (IO) redirections to the FDE. 
\item PDE finally encrypts or decrypts data with the \texttt{dm-crypt} kernel module and communicates with the physical volume and commits data on the physical volume or reads data from it.
\end{enumerate*}

\subsubsection{User steps}
Steps for using MobiGyges:
\begin{enumerate}
\item Boot the device with MobiGyges. The outer volume is mounted automatically. After booting up the device, the user can start using it as an ordinary device for daily purposes.
\item When using hidden volumes, open TriggerApp and enter a level (level of deniability is detailed in {\refs{sec:composition}}) of the password to mount the corresponding hidden volume without rebooting the device.
\item If an attacker suspects that PDE exists on the device and conducts the fill-to-full attack, the user can use level 0 before then (detailed in {\refs{sec:functionality}\ref{sec:fill_to_full}}) to defend the attack.
\end{enumerate}

  \begin{figure}[t]
\centering
\includegraphics[width=0.8\columnwidth]{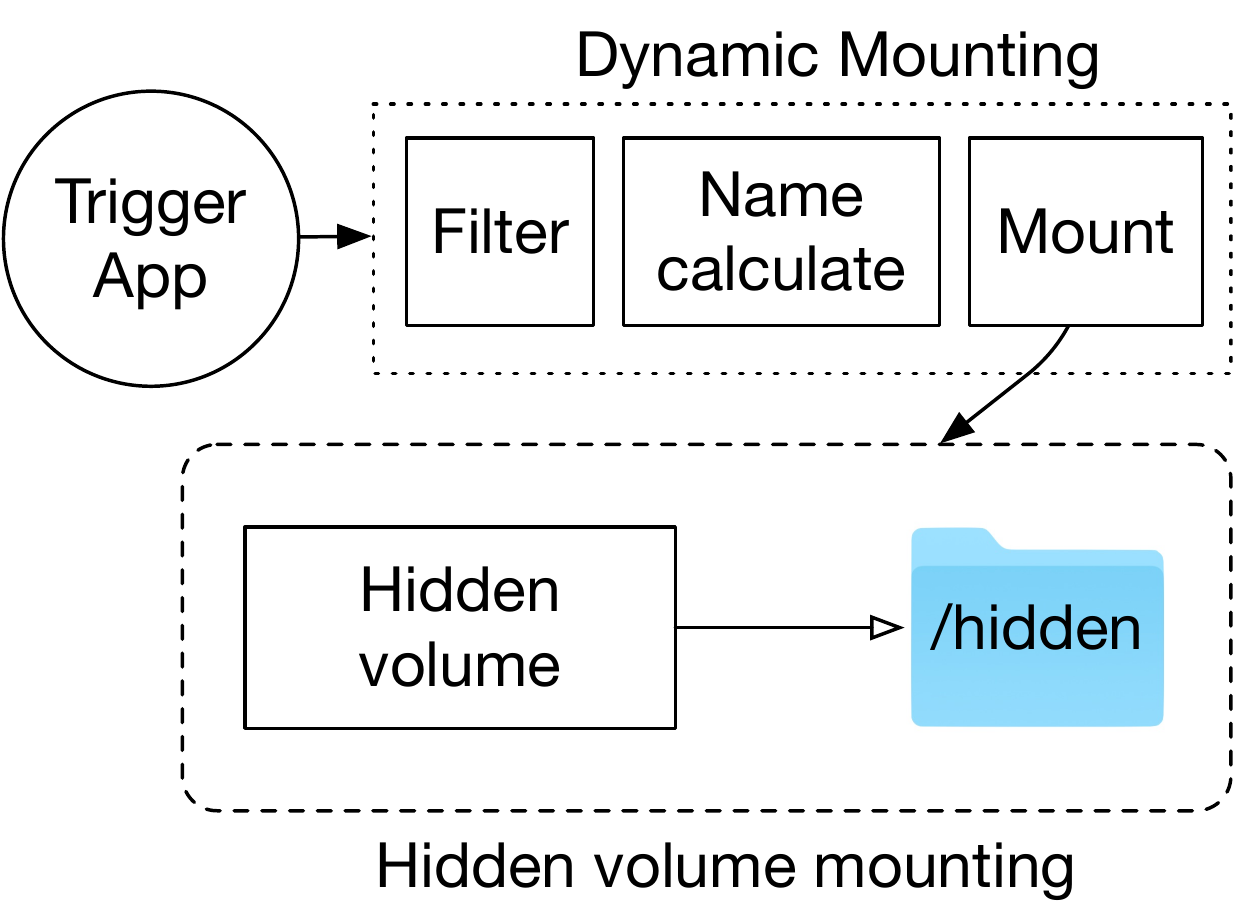}
\caption{The Dynamic Mounting service structure. Authorized App can mount the hidden volume on-demand.}
\label{fig:dynamic_mounting}
\end{figure}


\subsection{The Volume Management Module}
\label{sec:component}

This subsection details the design of the key part of MobiGyges, the Volume Management module. The Volume Management module leverages \textit{Thin Provisioning} and \textit{Device Mapper} by converting the physical storage into a Thin Pool and using the virtual logical volume to manage the physical storage in a fine-grained manner. In the rest of this subsection, we show how fine-grained storage blocks used by the Volume Management module solves  the aforementioned problems and defends the attacks, and we then anatomize each services in the module.

\begin{enumerate}[wide,label=\textbf{(\arabic*)}]

\subsubsection{Volume Management Module Circumvention}
\label{sec:functionality}

\item \textbf{Data Loss and Low Storage Utilization Elimination}

MobiGyges avoids data loss and improves the low storage utilization using the Volume Management module with the following steps as shown in \reff{fig:prob_avoid}.

\begin{enumerate*}[label=\roman*)]
\item The Volume Management module splits the whole physical disk into fine-grained storage blocks (i.e., 64KB), and then,
\item the module allocates each fine-grained storage block only for one volume, so the storage space of each volume is independent, and the data on each fine-grained storage block cannot be override by other volumes. Hence, the data loss problem is avoided.
\item The module tracks the usage of each fine-grained block, and stores volumes and the storage block mapping information (metadata) in a special format and encrypts the metadata with FDE on the physical disk. The metadata is stored in the metadata logical volume as shown in \reff{fig:thinpool_vlv}. When allocating storage space for the metadata logical volume, the available storage blocks are first sorted from big to small, and the least storage block that is larger than the needed storage size is used. In the figure, we can see that 
\item no ``reserved area'' is used because of the exclusive access of the same storage block from volumes, and hence the storage utilization is improved.
\end{enumerate*}

\begin{figure}[t]
\centering
\includegraphics[width=\columnwidth]{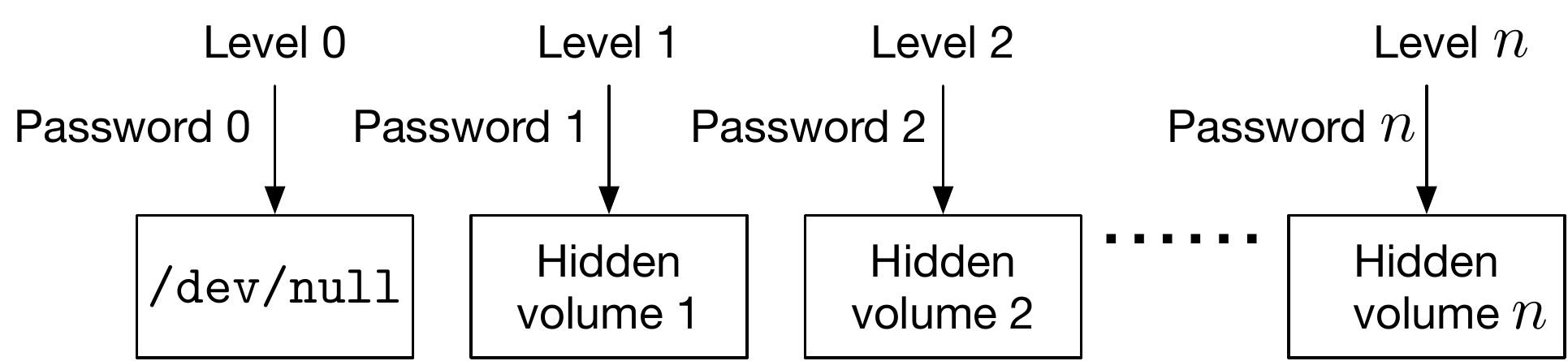}
\caption{Multi-level deniability. Level 0 is used to defend the fill-to-full attack.}
\label{fig:multi_level}
\end{figure}


\item \textbf{Device Reboot Avoidance}

MobiGyges avoids device reboot when using the hidden volume by leveraging the Dynamic Mounting service in the Volume Management module. The Dynamic Mounting service first filters the mounting request by identifying the owner of the request, and it only allows authorized applications (e.g., the TriggerApp) to mount the hidden volume. The mounting request is sent associated with a token generated by using OAuth \cite{oauth}, and thus if the token is not valid, the Dynamic Mounting service rejects the request. Then the Dynamic Mounting service calculates the name of the hidden volume based on the given PDE password and mounts the hidden volume on the mounting point. The procedure is shown in \reff{fig:dynamic_mounting}.

\item \textbf{Capacity Comparison Attack Defense}
\label{sec:cap_comp}

The attacker may conduct the capacity comparison attack and the fill-to-full attack to expose the hidden volume and compromise the sensitive data.

MobiGyges leverages the \textit{Shrunk U-disk method} to defend the capacity comparison attack. Shrunk U-disk is a kind of disk labeled with a bigger capacity, and operating systems also display the labeled capacity to users. However, the real capacity of the disk is smaller than the labeled capacity. For example, a memory stick is labeled to have 32GB storage, and the operating system also shows the capacity is approximately\footnote{It is a convention that storage hardware vendors using 1k=1000 rather than 1k=1024 to calculate the capacity.} equal to 32GB, but its actual capacity is only 8GB. Only when exhausting more than 8GB storage space on it, can the user find out the trick. However, users usually would not do that and cannot discover the ``trick''. We can exploit this ``trick'' to defend the capacity comparison attack by simply modifying the capacity of the outer volume to be the same as the physical capacity, which protects the hidden volume. The implementation choice of Shrunk U-disk method is detailed in Section {\ref{sec:composition}}. 

The challenge is how we can modify the volume capacity. We leverage \textit{Thin Provisioning} to fill the gap. Thin Provisioning is a novel technology designed to address the storage waste problem in the Cloud data center (DC). Because the servers are usually installed with larger physical storage capacity than the actual size of data usage for future demands. This is called \textit{Thick Provisioning}, but the physical storage cannot reach its capacity upper limit before replacing the disk with a larger one. Hence, a large amount of storage space is wasted. Thin Provision allows operators to flexibly preconfigure a bigger capacity of virtual volumes than the physical capacity, and physical storage spaces are not be used until data commit on them. When the physical storage exhausts, the operator can install new physical disks without modifying any previous settings. We employ the flexibility of Thin Provisioning and set the capacity of the outer volume to be the physical capacity to mitigate the capacity comparison attack.

\item \textbf{Fill-to-Full Attack Defense}
\label{sec:fill_to_full}

It is natural to conduct the fill-to-full attack when the attacker suspects the existence of the hidden volume(s). We leverage \textit{multi-level deniability} to defend the attack. Multi-level deniability allows users to define different levels of importance of sensitive data, and each level of sensitive data is associated with one hidden volume. To defend the fill-to-full attack, we introduce a special level (level 0) of deniability, which is associated with \texttt{/dev/null}. As shown in {\reff{fig:multi_level}}, when the attack discovers the hidden volume and tries to conduct the fill-to-full attack, the user uses the level 0 deniability and gives the device to the attacker to perform the fill-to-full attack. Attack data will first be written to the outer volume and recorded by the Volume Management module. When the physical disk exhausts, new attack data will be directed to \texttt{/dev/null}, and the Volume Management module will occur a full storage error when the recorded size of attack data plus the used capacity of the outer volume before the attack equals physical capacity.

\end{enumerate}

\subsubsection{Volume Management Module Composition}
\label{sec:composition}
In this subsection, we bring the technical design details of the Volume Management module and present how the Volume Management module manages the physical disk.

\begin{figure}[t]
\centering
\includegraphics[width=0.9\columnwidth]{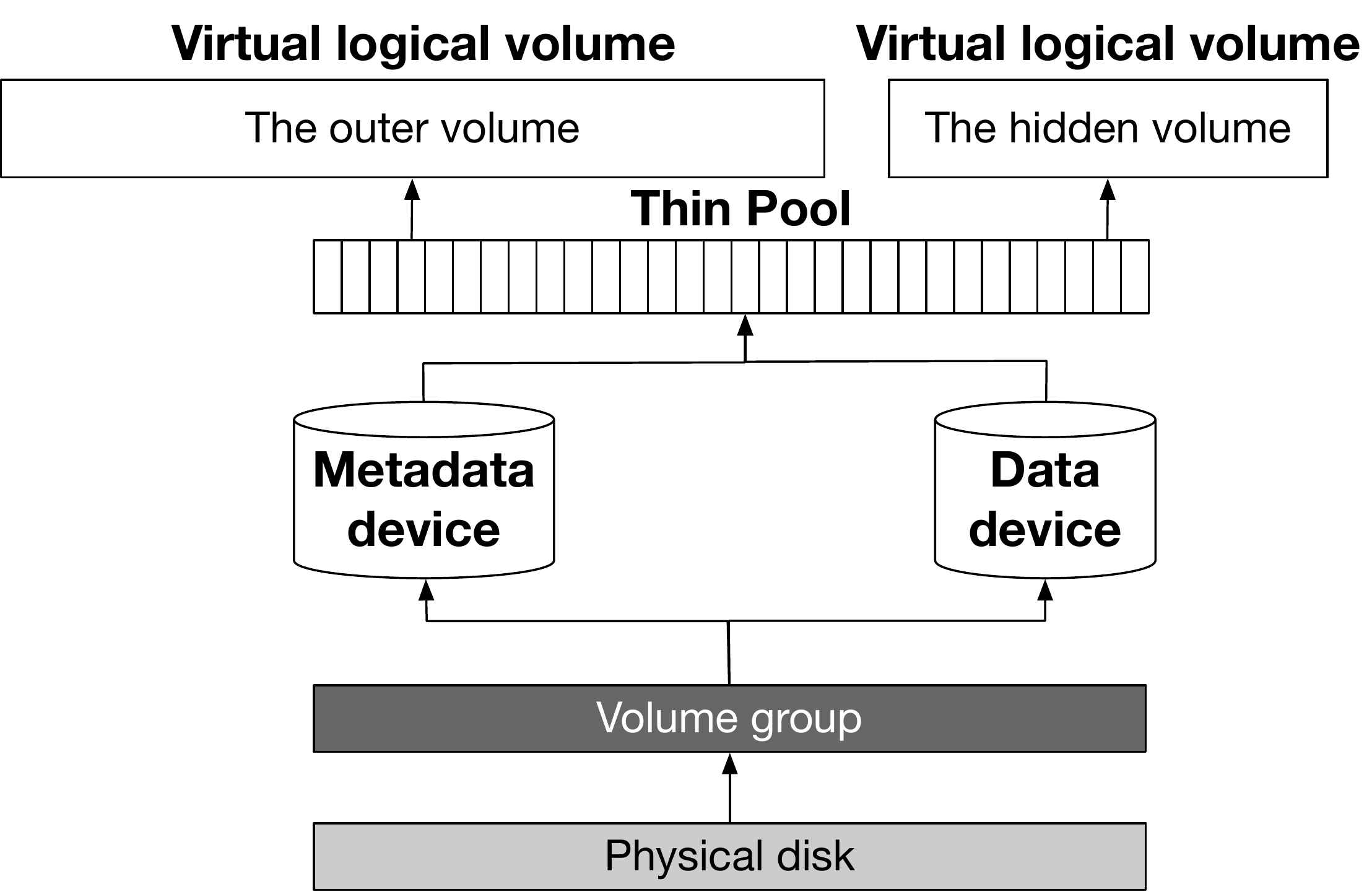}
\caption{The creation of the outer volume and hidden volume. The outer volume and the hidden volume are virtual logical volumes.}
\label{fig:thinpool_vlv}
\end{figure}


The Volume Management module leverages \textit{Thin Provisioning} and \textit{Device Mapper} to convert the physical disk into a \textit{Thin Pool}, which is a resource pool consists of fine-grained storage blocks. The Volume Management module then create \textit{virtual logical volumes} as the outer volume and hidden volumes. The storage spaces of the virtual logical volumes are allocated from the Thin Pool, and each virtual logical volumes allocates storage blocks distinctly from others.As depicted in {\reff{fig:thinpool_vlv}},
\begin{enumerate*}[label=\roman*)]
\item the Volume Management module converts physical disk into a volume group with Device Mapper. Each volume group consists of one or more physical disks.

\item Then, the Volume Management module creates two logical devices using the created volume group with  Device Mapper, namely, the data device and the metadata device. The data device is the storage space of Thin Pool, and the metadata device records the usage of each fine-grained storage block. Device mapper is used to map existing block devices into another logical block device. Device mapper redirects or filters IO requests from logical block devices to the mapped device (physical disk in our case). The process can be formalized by a 5-tuple,
$\langle O_{l}, S_{l}, T, D_{p}, O_{p}\rangle$,
where $O_{l}$ denotes the sector offset of mapping logical volume block, and $S_{l}$ is the number of sectors of the original volume block. $T$ denotes the type used to describe the way of mapping. $D_{p}$ denotes the mapped device, and $O_{p}$ denotes the offset of mapped physical volume block.

\item Next, the Volume Management module converts the the two devices into a Thin Pool.

\item Finally, the Volume Management module creates two virtual logical volume by using fine-grained storage blocks in Thin Pool with Device Mapper.

\end{enumerate*}

We anatomize each components in the Volume Management module in the rest of this subsection.

\begin{enumerate}[label=\textbf{(\arabic*)}, wide]

\item \textbf{Thin Pool} is the resource pool used for virtual logical volumes and created atop the encrypted logical volume. As shown in \reff{fig:layout}, it virtualizes the encrypt logical volume as a resource pool by carefully splitting the whole storage space into fine-grained storage blocks, and the outer volume and the hidden volumes allocate storage space from the pool on demand. When one storage block in the Thin Pool is allocated by one virtual logical volume, the storage block will be marked as used, and it will not be allocated for other virtual logical volumes. Hence the overriding problem is addressed. The mapping of fine-grained storage blocks and virtual logical volumes is stored in the metadata device in \reff{fig:thinpool_vlv}. In our case, each block is 64KB in size. Since each block is exclusively used by the hidden volume or the outer volume separately, the ``reserved area'' is needless. Thus, the storage utilization also improves.

\begin{figure}[t]
\centering
\includegraphics[width=0.8\columnwidth]{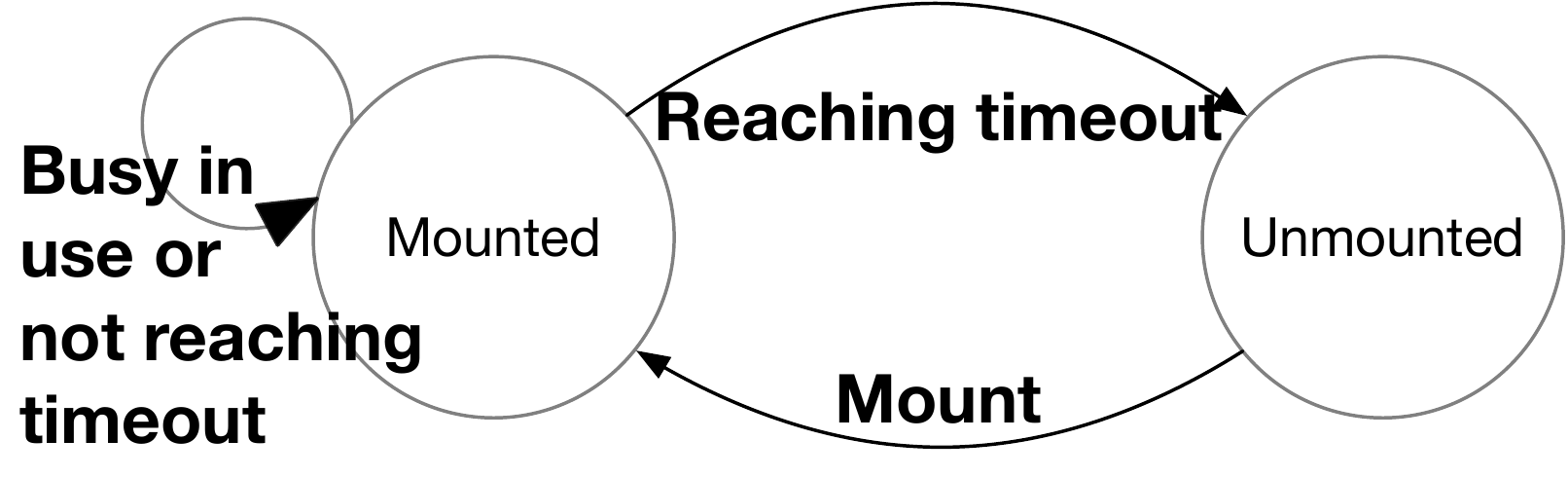}
\caption{Dynamic mounting state management.}
\label{fig:mount_state}
\end{figure}


\begin{figure*}[t]
\centering
\includegraphics[width=0.7\textwidth]{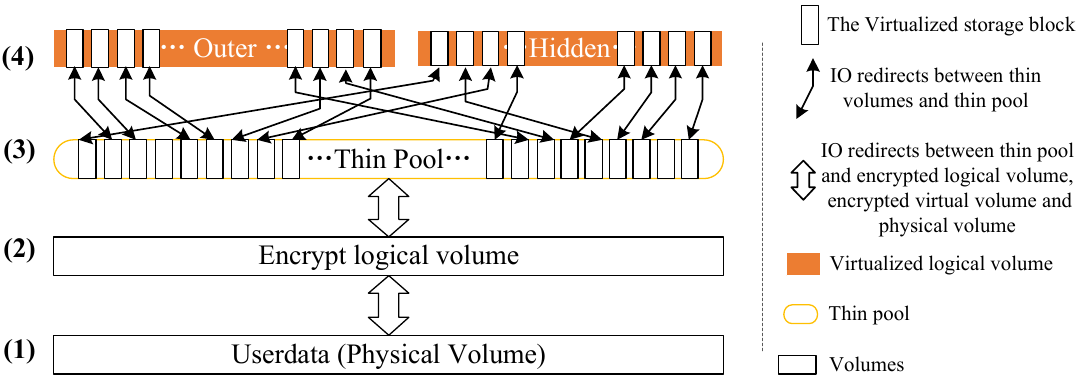}
\caption{Disk layout of MobiGyges. \textbf{(1)}~The \texttt{userdata} is the physical volume that used for storing user-generated data. \textbf{(2)}~FDE is applied to encrypt both data and the structure on top of it. \textbf{(3)}~all the \texttt{userdata} storage is virtualized as a Thin Pool. \textbf{(4)}~virtual logical volumes are created as outer volume and hidden volume, respectively.}
\label{fig:layout}
\end{figure*}


When creating the data and metadata logical volumes, the relationship between the size of the metadata volume and data volume can be formalized as follows.
\begin{equation}
\label{eq:meta_size}
S_m=\frac{S_p}{S_c}\times 64,
\end{equation}
where $S_m$ denotes the size of metadata volume. $S_p$ denotes the size of pool volume, and $S_c$ denotes the chunk size of pool volume.

\item\textbf{Virtual logical volume} is the final logical volume used as an outer volume or a hidden volume. The reason it is called ``virtual'' is when creating the virtual logical volume, its storage resources is not allocated until data commit on it. The capacity assigned at the creation time is only a label, and it does not indicate the actual capacity it possesses. The topmost part of \reff{fig:layout} shows the architectural level in the MobiGyges system. In MobiGyges, the capacity of the outer volume is configured as big as the physical volume to defend the capacity comparison attack. Device mapper is used to create the virtual logical volume from Thin Pool. For applications, the usage of the virtual logical volume has no differences between a physical volume.

%

\item\textbf{Multi-deniability}\label{sec:multideniability} allows users to define different levels of importance for their sensitive data and put them into different hidden volumes corresponding to different levels of deniability. MobiGyges provides multi-deniability by creating multiple hidden volumes. Each hidden volume is reserved for providing a specific level of deniability. To this end, even if the hidden volume is exposed, users may still deny the existence of sensitive data because the attacker fails to know the number of deniability the system provides, and which hidden volume is used to store the sensitive data. Each hidden volume has its name and password, and the name is calculated from the password. Without the password, the system cannot find the correct name of the hidden volume and cannot mount the hidden volume and thus fails to fetch the sensitive data stored on it. The name of each hidden volume is calculated with \refe{eq:hidden_name}.
\begin{equation}
\label{eq:hidden_name}
N = T(h(passwd + salt), b),
\end{equation}
where, $N$ is the final name. $T(s, b)$ is a trim function that trims $s$ into a $b$-length string. $h(x)$ is a hash function that hashes $x$ into a hash value. $passwd$ is the user password towards a hidden volume, and $salt$ is used for counter rainbow table attack~\cite{narayanan2005fast}.

\item\textbf{Dynamic Mounting} mounts the hidden volumes on demand. Unlike previous solutions, which requires rebooting the device and logging into the PDE mode to use the hidden volumes. MobiGyges can use hidden volume on demand needless device rebooting to switch to the PDE mode. Therefore, authenticated applications can use dynamic mounting to mount the needed hidden volume instantly. Due to the confidentiality of the hidden volume, operations of authenticated applications should be cautious. When requesting for mounting the hidden volume, an access token is needed to validate the application, and if the token is not valid, the mounting request is rejected.

Android is a UNIX-like system, and in
 a UNIX-like system, devices are presented as block device files. Each block device file has its own name. Therefore, when using a hidden volume, users first enter the password of the wanted hidden volume at authenticated application such as TriggerApp and MobiGyges will mount the correct hidden volume based on \refe{eq:hidden_name}. As depicted in \reff{fig:mount_state}, when the hidden volume is mounted, a timer is started, and the hidden volume will be automatically unmounted as the timer times out. This mechanism protects the hidden volume from being discovered by examining the currently mounted devices. 

\end{enumerate}

\subsubsection{TriggerApp}
\label{sub_sec:trigger}
TriggerApp is the interface among the users and MobiGyges's hidden volumes. Users have to use TriggerApp to trigger the special operations and dynamically mount the hidden volume for storing sensitive data. TriggerApp functionality should be secret to prevent MobiGyges from exposing. Therefore, TriggerApp is recommended to be implemented inside a system built-in application.

\section{Implementation}
\label{sec:impl}
We implement our MobiGyges prototype system on LineageOS 13 for Google Nexus 6P. We choose LineageOS rather than original Android Open Source Project because LineageOS provides necessary vendor specific hardware adaptation like camera and baseband driver supports. For the Volume Management component of MobiGyges, we add about 500 lines of C code to LineageOS, and we port Logical Volume Management (LVM) and Thin Provisioning tools (pdata\_tools) \cite{thinprovisioningtools} and some system building scripts to Android. For TriggerApp, we add approximately 500 lines of Java code. In this section, we present the implementation challenges and considerations of MobiGyges.

\begin{enumerate}[label=\textbf{(\arabic*)}, wide]

\item\textbf{Manipulating the \texttt{userdata} partition is hard on Android}. MobiGyges initialization requires mounting and unmounting the \texttt{userdata} partition. However, the \texttt{userdata} partition cannot be unmounted while the system is running because some system files are stored on the \texttt{userdata} partition, and these files are busy in use \cite{shao2014rootguard}. For system stability and data integrity, the operating system does not allow the \texttt{userdata} partition to be unmounted while busy. However, the creation of Thin Pool and virtual logical volume requires a free (not in used) partition. To this end, we put the Thin Pool and virtual logical volume creation procedure executing at the system booting stage before mounting the \texttt{userdata} partition. We put the code right after the FDE procedure in Android Volume Daemon (VOLD) located at \texttt{cryptfs.c}. The code executes \texttt{lvm} toolset by forking a child process.

\item\textbf{Running toolsets on Android} is another challenge for us because the toolsets are usually for desktop and server that are x86 platforms rather than the ARM-based mobile platforms. Luckily, Android is based on Linux kernel and has the kernel modules needed by Thin Provisioning and Device Mapper, and the system calls are the same as desktop and server versions. Therefore, porting LVM and pdata\_tools that are used to build logical volumes, Thin Pool, and virtual logical volumes does not require much code and building system modification. We use \texttt{gcc-arm-linux-androideabi} to conduct the cross-compile. We first intended to compile the source directly, but these tools require libraries that should be cross compiled in advance and made them Android runnable. Apart from that, \texttt{-enable-static\_link} and \texttt{LDFLAGS=-static} flags should be set to ensure the compilation target is statically executable. Otherwise, the toolsets cannot run when pushing to Android because Android lacks of these libraries. We also modified the code in the Android build system to compile the toolsets into the final installation package.

\item\textbf{TriggerApp implementation} is an important part of the work of implementing MobiGyges. We implement our TriggerApp into the Android system built-in Calculator. Apart from the basic calculation functionality, our tailored Calculator has the extra functionality like secret recording, filming, and picturing. The calculator could still make calculations, and otherwise, the system is prone to expose the special design and thus compromise sensitive data. As we all know, dividing any number by 0 is an illegal operation. When users attempt to do such an illegal operation, the system will throw an exception and prompt an alert to users to indicate the calculation is not allowed. Thus, we change the exception processing behavior of the calculator by modifying it to display a special operation Use Interface (UI) according to what the user enters. Identifying the divided by 0 operation can be done by parsing the input \texttt{String} every time when the user presses the '=' button. However, users usually just conducts normal calculations operations, and the input parsing is worthless under such condition. Android applications are written in Java\footnote{C/C++ can also be applied to Android application to improve the running efficiency}. Java has an \texttt{ArithmeticException} that whenever an illegal operation occurs, it throws an \texttt{ArithmeticException}. We put our parsing code into the \texttt{catch \{\}} code block. When an \texttt{ArithmeticException} is triggered, the code will analyze the input and process the operations accordingly.

\item\textbf{Dynamic mounting} is implemented as SystemService. Normally, all kinds of Android supporting hardware has the SystemServices and Hardware Abstract Layer (HAL) definition pair. Apart from that, lightweight Service and lightweight HAL are also provided as pairs in Android. For example, the WiFi module has both Service and HAL definitions. But the HAL definition is needless in our system because no new physical hardware devices are added to the system. We implement the MobiGyges SystemService by using the method provided by~\cite{yaghmour2013embedded}. Inside the dynamic mounting SystemService, it 
\begin{enumerate*}[label=\roman*)]
\item calculates the trimmed hash value with the input password and salt by using \refe{eq:hidden_name}. The salt, uses the crypto footer of the \texttt{userdata} partition in our case, and 
\item it mounts the corresponding hidden volume with the calculated device name and the access token.
\end{enumerate*}
If either the password or the access token is incorrect, the mounting request is rejected.

\item\textbf{Full Disk Encryption (FDE)} is used to protect the Volume Management module. MobiGyges first performs FDE by creating an encryption layer on the \texttt{userdata} partition, which makes it easy to encrypt all data on the \texttt{userdata} partition. MobiGyges uses 128-bit Advanced Encryption Standard (AES) \cite{miller2009advanced} with cipher-block chaining (CBC) and ESSIV:SHA256 to perform the encryption. The master key is encrypted with 128-bit AES via invocations to the OpenSSL library. It is recommended that users use 128 bits or more for the key (with 256 being optional) to improve the security. \reff{fig:layout}\textbf{(2)} represents the encrypted logical volume created atop the physical volume. It also shows the logical position in MobiGyges. FDE is created using the Device Mapper technology. 

\end{enumerate}

\section{Evaluation}
\label{sec:evalueate}

In this section, we describe experiments on storage utilization and performance overheads by comparing MobiGyges with state-of-the-art works. We conduct experiments on Google Nexus 6P mobile phone with 3GB LPDDR4 DRAM and an octa-core CPU with 4 Cortex-A53@1.55GHz cores and 4 Cortex-A57@2GHz cores. We use \texttt{dd}~\cite{dd}, bonnie++~\cite{coker2012bonnie++}, and AndroBench~\cite{kim2012androbench} to conduct the experiments.

\subsection{Performance Evaluation Tools}
\label{sec:perf_tools}

\texttt{dd} copies blocks of data from one file to another and is provided by most UNIX platforms. It allows parameters like r/w buffer size that can be easily used for IO performance study. We also use the \texttt{fsync} parameter in \texttt{dd} because it does not bypass the kernel disk caches, and when it writes data to the device, the data may still not be committed on the device upon \texttt{dd} completion. 

We use Bonnie++ in our experiments: an IO benchmark tool suite that aims to perform several simple tests of hard drive and file system performance. It has 2 types of tests. The first is to test the IO throughput. The second is to test creation, reading, and deleting operations on many small files. We use the second type in our test.

AndroBench is a popular benchmark tool for Android. It is an Android application, which provides sequential/random r/w tests and SQLite benchmarks. Since Android uses SQLite as its built-in database for applications, the performance of SQLite is trivial for Android Apps user experience and is closely correlated with the storage performance.

\begin{figure}[!t]
\centering
\includegraphics[width=0.9\columnwidth]{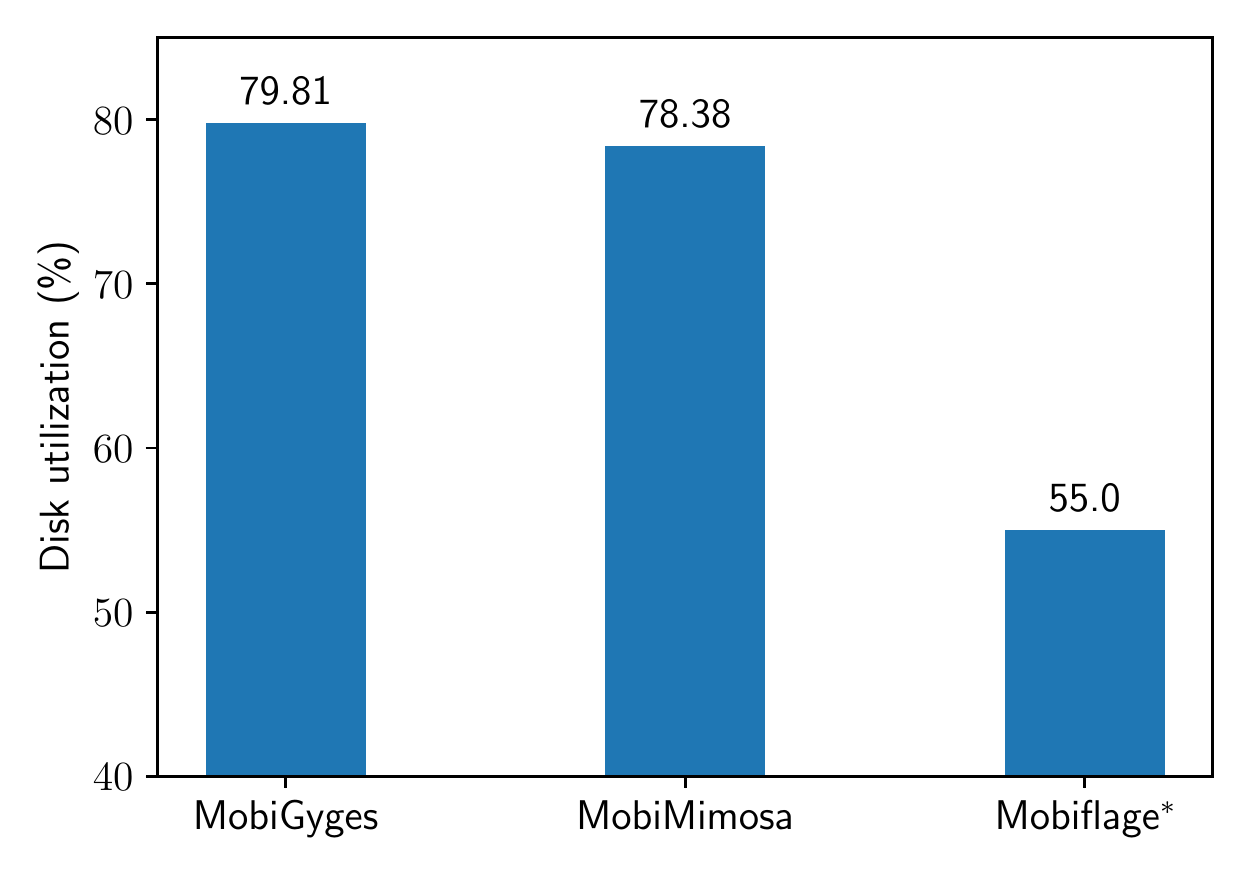}
\caption{Disk utilization result of MobiGyges (1 hidden volume) and Mobiflage. Mobiflage$^*$ represents all the systems uses the same mechanism as Mobiflage.}
\label{fig:util}
\end{figure}


\subsection{Storage Utilization Evaluation}
\label{sub_sec:storage}

MobiGyges creates the Thin Pool using \refe{eq:meta_size} to calculate the size of metadata volume. Hence, the capacity of Thin Pool equals to the size of the data volume, so the size of the data volume is the actual capacity that allows users to store their data on the device. To this end, physical storage utilization $\eta$ can be calculated with \refe{eq:utilization}. The definitions of the notations used in \refe{eq:utilization} are the same as that of \refe{eq:meta_size} that can be found in \reft{tbl:notation}.
\begin{equation}
\label{eq:utilization}
\begin{aligned}
\eta &=\frac{S_p}{S_p + S_m}\\
    &=\frac{S_p}{\frac{S_p}{S_c} \times 64 + S_p}  \times 100\% = 99.9024\%.
\end{aligned}
\end{equation}


We compare disk utilization of Mobiflage and MobiMimosa with MobiGyges because most of the related state-of-the-art works \cite{chang2015mobipluto, userfriendly, mobiceal, yu2014mobihydra} use the same method as Mobiflage. For all experiments, we use \texttt{dd} to conduct experiments, and each has 50 trials\footnote{Results are stable in the first 20 trials, and we conduct another 30 trials to decrease the errors.}. 

As shown in \reff{fig:util}, the average disk utilization of MobiGyges is 79.81\%. The 20\% loss is mainly because each Ext4 file system takes up about 10\% of the storage for its metadata use~\cite{mathur2007new, Chen:2017:SUE:3072970.2820488}, and we have two volumes formatted with Ext4, one for the outer volume and the other for the hidden volume. Therefore, the storage space utilization of MobiGyges is 99.76\%, which is equivalent to the theoretical result shown in \refe{eq:utilization}.
\begin{equation}
\label{eq:mobiflage}
\begin{aligned}
\mbox{\textit{Offset}} &= \lfloor 0.75 \times \mbox{\textit{vlen}} \rfloor \\
&- (H(\mbox{\textit{pwd}} || \mbox{\textit{salt}})\mbox{mod} \lfloor 0.25 \times \mbox{\textit{vlen}}\rfloor).
\end{aligned}
\end{equation}

Mobiflage \cite{skillen2014mobiflage} proposes \refe{eq:mobiflage} to calculate the offset of placing the hidden volume, and this equation indicates the capacity of the hidden volume is between 25-50\% of the total disk capacity. Therefore, Mobiflage introduces up to 25\% of the storage waste. Since each file system takes another 10\% of the capacity for metadata, and there are two file system instances used (one used for the outer volume and the other used for the hidden volume), there is 20\% of inevitable waste on each device. Thus, the total storage waste for Mobiflage is 45\%. We want to get the improvement purely benefitting from our MobiGyges design, so we add 20\% for each of the experimental results, and hence, MobiGyges increases the storage utilization by over 30\%\footnote{$((79.81\% + 20\%) - (55\%+20\%)) / 55\%+20\%)\approx 30\%$}. 


MobiMimosa \cite{hong2017personal} achieves storage utilization improvement by using the \texttt{dm\_table} to record the storage block used by the hidden volume for the outer volume. Therefore, the only waste is the \texttt{dm\_table}. MobiMimosa uses the Ext4 file system, and the default block size of the Ext4 file system on Android is 4KB. Consequently, if a hidden volume has 5GB capacity, the size of the \texttt{dm\_table} file can be up to 64MB\footnote{Each \texttt{dm\_table} item is 80 byte.}, which means the size of \texttt{dm\_table} is 1.25\% of the capacity of the hidden volume. Since the maximum size of a hidden volume is usually smaller than a half of the total physical disk capacity, the size of the \texttt{dm\_table} file can take up to 0.625\% of the total physical storage space. We test the storage utilization on Google Nexus 6P, with the same setting as MobiGyges. The results show that the overall storage utilization of MobiMimosa that has one 5GB hidden volume is 78.375\%.


\subsection{Performance Overhead Evaluation}
\label{sub_sec:performance}

Performance overhead on the PDE system is critical because the attack may compromise the PDE system if the performance overhead is significant. We evaluate the IO performance overhead of MobiGyges (Outer and Hidden in the figures) by comparing the IO performances of MobiGyges with that of Android FDE (Baseline in the figures) because Android FDE is enabled by default, and MobiGyges uses Android FDE as a foundation. In the rest of this subsection, we use different IO benchmark tools to evaluate the performance of the system and analyze the overhead.


\begin{figure}[!t]
\centering
\includegraphics[width=0.9\columnwidth]{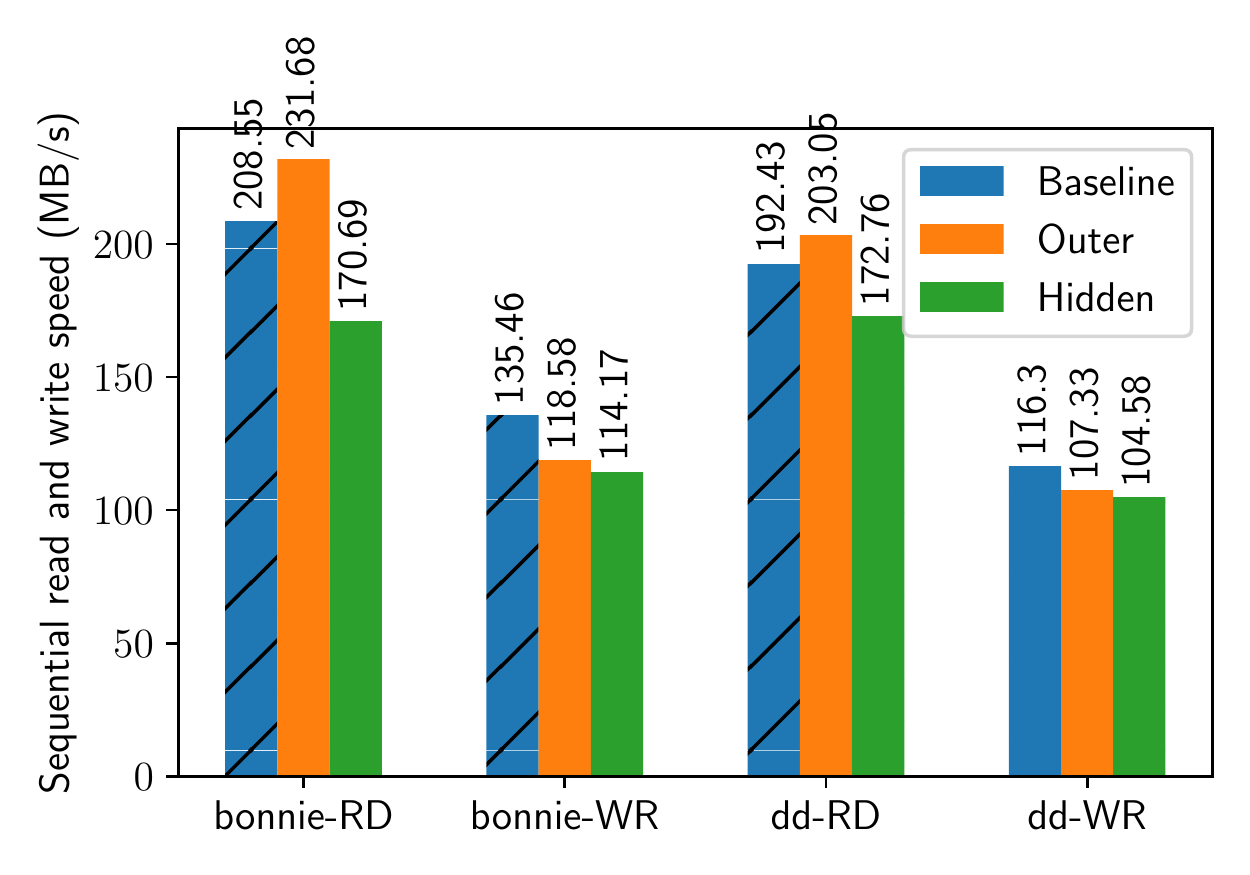}
\caption{Bonnie++ and \texttt{dd} IO performance test. \textbf{Baseline} is the original Android FDE. \textbf{Outer} is the MobiGyges's outer volume, and \textbf{Hidden} is the MobiGyges's hidden volume.}
\label{fig:bonnie_rdwr}
\end{figure}


\begin{figure*}[!t]
\centering
  \includegraphics[width=0.8\textwidth]{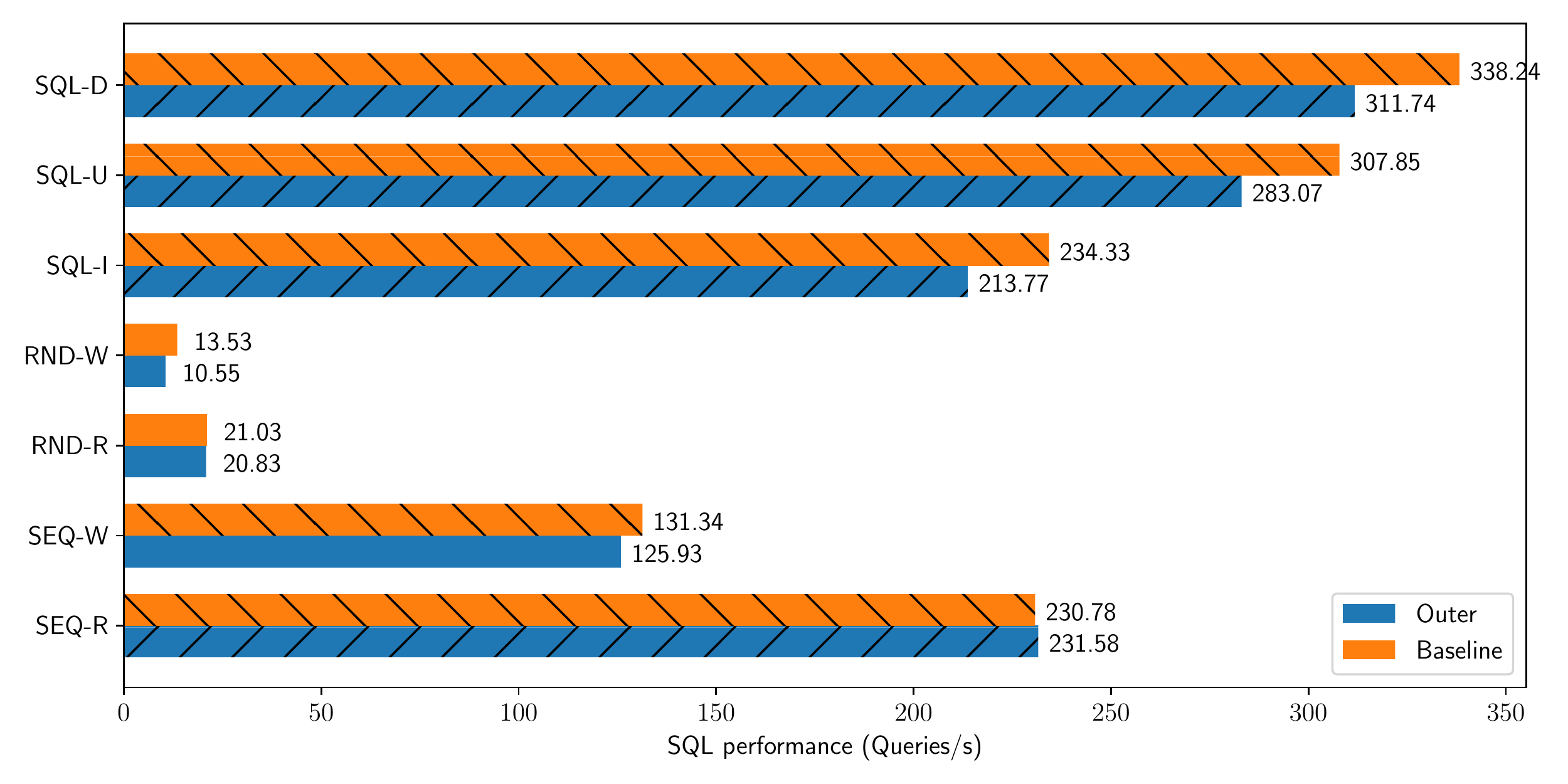}
  \caption{Sequential read and write speed and random read and write speed comparing between outer volume and original Android FDE. Tested with AndroBench. The unit of SQL operations is Q/s (query per seconds). \textbf{Outer} is the MobiGyges's outer volume and \textbf{Baseline} is the original Android FDE.}
  \label{fig:androbench_rdwr}
\end{figure*}


\textbf{(1) Bonnie++ test:} We run 50 trials\footnote{The results are very stable, we run 50 trials to minimize the error.} of sequential block tests on a 6GB file\footnote{Bonnie++ requires to test on a file whose size is twice as big as the device RAM to decrease the influence of system cache.} on each system. \reff{fig:bonnie_rdwr} shows that the outer volume outperforms approximately 10\% over Android FDE in terms of reading.  The reason for MobiGyges's outer volume performs better is because of the IO request batching mechanism. When reading sequentially, the system can merge different IO requests, cache them, and read a bunch of adjacent storage blocks together~\cite{tpkernel}. However, the write performance is reduced by about 12\%. We analyze the impact of the performance overhead by considering the user experience. With the same 1GB file, we calculate the time difference between MobiGyges and the baseline (Android FDE). Therefore, in terms of reading, MobiGyges reduces about 500ms compared with the original Android FDE. Similarly, MobiGyges takes 600ms longer than the original Android FDE. We believe the performance penalty is acceptable. Comparing the hidden volume and outer volume of MobiGyges, the write performance is only reduced by 3\%. The reduction results from the extra encryption over the hidden volume.

\textbf{(2) \texttt{dd} test:} We use the \texttt{dd} command to generate data from \texttt{/dev/zero} and write data on the volume with buffer size 600MB for once\footnote{\texttt{dd if=/dev/zero of=disk.img	bs=600M count=1 conv=fsync}} to test writing performance of MobiGyges and the original Android system. Then, we use \texttt{dd} to read data from the volume and write the output to the \texttt{/dev/null} "blackhole"\footnote{\texttt{dd if=disk.img of=/dev/null bs=600M count=1}}, to test the reading performance of MobiGyges and the original Android system.  Normal \texttt{dd} version prints the statistical data after each command finishes. However, Android system uses the busybox~\cite{busybox} version \texttt{dd} tool that prints nothing after completion. Thus, we use \texttt{time} command to count the execution time. Moreover, we clear the cache in RAM before every tests\footnote{\texttt{echo 3 > /proc/sys/vm/drop\_caches}}, which eliminates the error caused by operating system caching mechanism. As shown in \reff{fig:bonnie_rdwr}, \texttt{dd} shows equivalent results as that of bonnie++. 


\textbf{(3) AndroBench test:}  Due to the permission control of AndroBench, we fail to test the performance of the hidden volume, but comparing the performance differences between the Android FDE and MobiGyges's outer volume is still sufficient to show the performance overhead of MobiGyges.
\reff{fig:androbench_rdwr} depicts the result of  the AndroBench tests. In the random r/w (RND-W and RND-R) and sequential write (SEQ-W) tests, MobiGyges has a performance penalty due to the Thin Pool and device mapping. In the sequential read (SEQ-R) test, MobiGyges outperforms the baseline because of merged IO requests and batch fetching mechanism. For SQL queries, MobiGyges reduces about 8\% of the performance compared with the baseline. However, in terms of modifying data, which writes data on the volume, the performance penalty can be up to 22\%. SQLite uses VDBE (Virtual Database Engine) as its background engine, and VDBE uses the B-Tree data structure to store data on a file system. VDBE also adopts the concept of paging as a unit to allocate space for the value of a key, which is similar to pages in the operating system virtual memory mechanism.  The size of a page is fixed, so if the record is bigger than a page, it has to be stored into several pages that are linked together using pointers. To this end, it has to first read all the pages associated with the key before VDBE can finally update the record. Consequently, the multiple writes downgrade the performance of SQL queries, but We believe the  performance penalty for the SQL operations is acceptable~\cite{1607949}.

\subsection{Performance Overhead Comparison}
\label{sec:overhead_comparison}

In this subsection, we compare the performance overhead between MobiGyges and some recent related works. Note that MobiGyges is not optimized for improving the IO performance. We post our IO performance overhead comparison experimental results here for the further optimization target.

\begin{figure}[!t]
\centering
  \includegraphics[width=0.9\columnwidth]{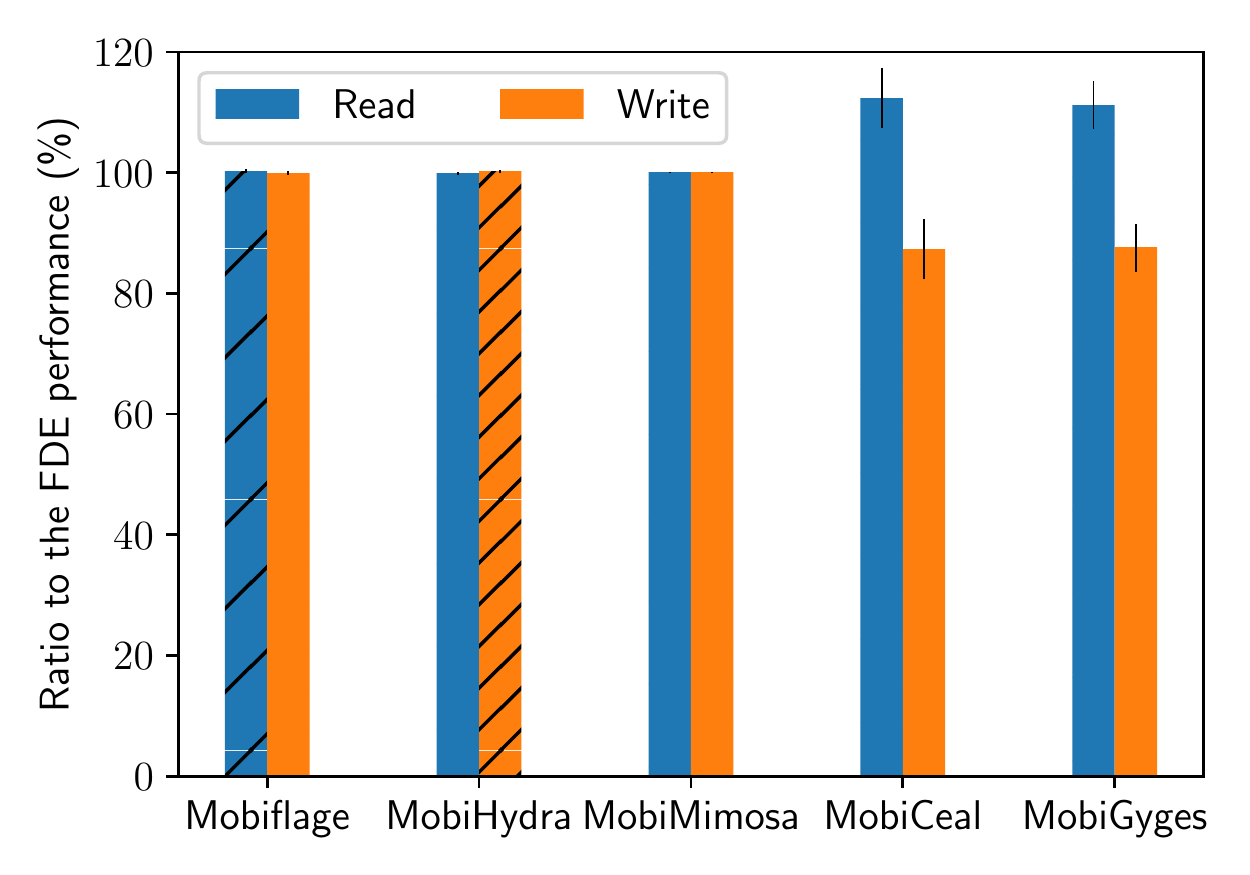}
  \caption{Performance overhead. the lower, the better. We use FDE (LUKS \cite{fruhwirth2009luks} on Linux) as the baseline.}
  \label{fig:system_comparison}
\end{figure}

We implement disk management part of Mobiflage~\cite{skillen2014mobiflage}, MobiHydra~\cite{yu2014mobihydra}, MobiMimosa~\cite{hong2017personal}, MobiCeal~\cite{mobiceal} on Linux desktop PC and evaluate their performance with \texttt{dd}. The experimental results are shown in \reff{fig:system_comparison}, and the baseline system is the Android FDE\footnote{We implement Android FDE by using LUKS on Linux.}. Mobiflage, MobiHydra and MobiMimosa are native solutions, and their outer volume and the hidden volume are implemented in the same way as Android FDE. Hence, they have the equivalent performance to the Android FDE. MobiCeal and MobiGyges leverage storage virtualization mechanisms, which introduce performance overheads. MobiGyges outperforms MobiCeal on the performance overhead.
%
%

\section{Discussion}
\label{sec:discussion}

In this section, we discuss counter-measurements of MobiGyges towards defending \textit{common attacks}, and we then discuss the drawback and possible future works.

\subsection{Security Discussion}

Common attacks are identified by the existing literature and can be solved by existing counter-measurements. MobiGyge's mechanism of defending these attacks makes no difference to existing ones, and the discussion is to show that MobiGyges considers defending common attacks by design. Experiments of validating these mechanisms have been listed as part of our future works.

\begin{enumerate}[label=\textbf{(\arabic*)},wide]
\item \textbf{Password guessing} is the attack by trying all possible password characters repeatedly in a brute force way to identify the correct password. MobiGyges provides both salt and retrial timeout mechanism that can efficiently defend the password guessing attack and the rainbow table attack.

\item \textbf{Raw data parsing} is conducted by parsing the physical disk raw data and try to reconstruct the files. MobiGyges uses FDE to defend the raw data parsing attack, in which all data are encrypted before committing to the physical disk.

\item \textbf{Encryption primitive leakage} means the type of encryption or data protection is leaked. In the hidden volume based PDE, it refers to hidden volumes are exposed and is generally conducted from parsing the raw physical disk data. MobiGyges allocates fine-grained storage blocks from Thin Pool for the outer volume and hidden volumes on demands, which stores data in a striped way. Thus, attackers cannot tell the belonging of each data block and fail to distinguish from volume to volume. Moreover, MobiGyges applies FDE for Thin Pool, which increases the complexity and protects hidden volumes.

\item \textbf{Flash storage leakage} refers to the NADN flash storage structure used by mobile devices could leak sensitive data because flash storage executes writing or wiping data in the unit of \textit{page}, and can only change some bits in the page to 0 or change all bits in the page to 1. Hence, writing happens only on an empty page (with all 1s). Thus, data needs another \textit{temporary} page for saving the current unchanged data in the original page. This temporary page can leak the sensitive data if not erased in time. MobiGyges addresses this problem with FDE, and data on flash pages is not the plain data but cipher data, which mitigates the flash storage leakage.

\item \textbf{Mobile carrier leakage} represents the possible inconsistent record from the mobile device and the mobile carrier provider. The existing literature separate working modes into two, and PDE can only be used under the PDE mode. At the time using the PDE mode, carrier informations (e.g., phone call history, cellular traffic usage) are stored on the hidden volume, and attackers will find that these informations recorded by the carrier provider are more than that recorded by the outer volume. Thus, this is prone to expose hidden volumes. MobiGyges eliminates mobile carrier leakage by removing the design of two modes, and all the informations are stored on the outer volume. Thus, carrier information records are consistent between the carrier provider and the outer volume, which mitigates the leakage.
\end{enumerate}

\subsection{Drawbacks and Future Works}

Although MobiGyges brings new hope to the PDE community, it falls short when the attacker can record the fill-to-full attack data and tries to retrieve the filled data from the device. One possible remedy is to employ data compression techniques to mitigate this issue. MobiGyges also opens up several interesting directions for future research. For example, theoretically evaluating the mathematically model of MobiGyges and existing works, reducing the performance overhead by replacing the Linux Thin Provisioning module with newly proposed high-performance Thin Provisioning tools (e.g., ThinStore~\mbox{\cite{thinstore}}), and conducting further experiments in terms of defending common attacks (e.g., rainbow table attack).

\section{Conclusion}
\label{sec:conclude}

This paper has presented MobiGyges, a PDE system that addresses the problems of data loss, storage waste, and device reboot on existing PDE systems by splitting the physical storage into fine-grained storage blocks, and each storage block is used only by one volume, and using Dynamic Mounting service to mount the hidden volume without rebooting the device. We have also identified the \textit{capacity comparison attack} and \textit{fill-to-full attack} targeted at PDE systems, and MobiGyges can jointly leverages the \textit{Shrunk U-disk method} and \textit{multi-level deniability} to defend them. We have implemented a proof-of-concept system on LineageOS~13 for real mobile devices. Experimental results show that MobiGyges achieves over 30\% storage utilization improvement with an acceptable performance overhead.

\section*{Acknowledgement}

This work was supported in part by the National Key Research and Development Program of China under Grant 2018YFB1003804, Natural Science Foundation of China under Grant 61921003, and the China Scholarship Council. We would also like to thank the editors and anonymous reviewers for their valuable comments and suggestions.

\bibliographystyle{elsarticle-num}
\bibliography{refs}



\end{document}